\documentclass[showpacs,twocolumn,showkeys,amsmath,amssymb,pra,superscriptaddress,nofootinbib]{revtex4-1}

\usepackage{color}

\usepackage{graphicx}   

\begin{document}

\title{Scattering bright solitons: quantum versus mean-field behavior} 

\author{Bettina Gertjerenken}
\email{b.gertjerenken@uni-oldenburg.de}
\affiliation{Institut f\"ur Physik, Carl von Ossietzky Universit\"at, D-26111 Oldenburg, Germany}

\author{Thomas P.\ Billam }
\affiliation{Department of Physics, Durham University, Durham DH1 3LE, United Kingdom}
\affiliation{Jack Dodd Center for Quantum Technology, Department of Physics, University of Otago, Dunedin 9016, New Zealand}

\author{Lev Khaykovich}
\affiliation{Department of Physics, Bar-Ilan University, Ramat-Gan, 52900 Israel}

\author{Christoph Weiss}
\affiliation{Institut f\"ur Physik, Carl von Ossietzky Universit\"at, D-26111 Oldenburg, Germany}
\affiliation{Department of Physics, Durham University, Durham DH1 3LE, United Kingdom}

                  
\date{03 September 2012}
 
\begin{abstract}

We investigate scattering bright solitons off a potential using both analytical and numerical methods. Our paper focuses on low kinetic energies for which differences between the mean-field description via the Gross-Pitaevskii equation (GPE) and the quantum behavior are particularly large. On the $N$-particle quantum level, adding an additional harmonic confinement leads to a simple signature to distinguish quantum superpositions from statistical mixtures. While the non-linear character of the GPE does not allow quantum superpositions, the splitting of GPE-solitons takes place only partially. When the potential strength is increased, the fraction of the soliton which is transmitted or reflected jumps non-continuously. We explain these jumps via energy-conservation and interpret them  as indications for quantum superpositions on the $N$-particle level.  { On the GPE-level, we also investigate the transition from this stepwise behavior to the continuous case.}
\end{abstract} 
\maketitle 


\section{Introduction}

Bright solitons generated from attractively interacting ultra-cold atoms have been realized experimentally both in quasi one-dimensional (1D) configurations~\cite{KhaykovichEtAl02,StreckerEtAl02} and three dimensions (3D)~\cite{CornishEtAl06}.

Some of the effects investigated for bright solitons~\cite{PethickSmith08} could, in principle, also be investigated in classical systems. However, bright quantum-matter-wave solitons are mesoscopic quantum objects which are particularly useful to investigate beyond mean-field effects of quantum solitons: While there are cases for which both the classical and quantum descriptions agree for particle numbers as low as $N\gtrapprox 3$~\cite{MazetsKurizki06}, it has been proposed to use scattering bright solitons off a potential in order to produce non-classical quantum superpositions~\cite{WeissCastin09,Streltsov09b}.

 Other investigations of bright solitons include stationary solutions of the {(mean field)} Gross-Pitaevskii equation (GPE)~\cite{CarrEtAl00b}, soliton trains~{\cite{AlKhawajaEtAl02,StreltsovEtAl11}}, solitons under transverse confinement~\cite{SalasnichEtAl02} {and incoherent matter-wave solitons~\cite{BuljanEtAl05}} as well as  deviation from one dimensionality~\cite{KhaykovichMalomed06} and regular and chaotic dynamics in soliton collisions~\cite{MartinEtAl07}. Research also covers topics ranging from {fragmented states~\cite{StreltsovEtAl08}},
 stabilization and destabilization of second-order solitons against perturbations~\cite{YanayEtAl09} and quantum reflections~\cite{CornishEtAl09} over soliton localization via disorder~\cite{KivsharEtAl90,Muller11} and in time-dependent traps with time-dependent scattering length~\cite{Khawaja09}, resonant trapping through a quantum well~\cite{ErnstBrand10}  and possible applications to interferometry~\cite{BillamEtAl11,MartinRuostekoski2012,HelmEtAl2012}.

The focus of the present paper lies on scattering bright solitons off a potential in a one-dimensional geometry. On the $N$-particle quantum level this can result in quantum superpositions; however, the detection of such superpositions requires the identification of a clear experimental signature. For mean-field solitons both behavior similar to the case of a single particle (see, e.g., Ref.~\cite{MartinRuostekoski2012,HelmEtAl2012}) and considerably deviating behavior has been reported in the literature.\footnote{{Scattering mean-field (GPE) solitons off a barrier with behavior that considerably deviates from a single particle has been investigated  both in Ref.~\cite{LeeBrand06} and on page~2 of Ref.~\cite{Streltsov09b}.}} By focusing on low kinetic energies, we will discuss the parameter regimes for which both types of behavior can be expected.

The paper is organized as follows: Section~\ref{sec:model} introduces the models used to investigate scattering bright solitons off scattering potentials. The energetically allowed final states are discussed in Sec.~\ref{sec:mesos}.

In Sec.~\ref{sec:scattering}, scattering bright solitons off a scattering potential is investigated  with additional harmonic confinement as used in experiment~\cite{Hulet10,*Hulet10b}.  Section~\ref{sec:bound} explains the behavior on the level of the GPE. Section~\ref{sec:concl} concludes the paper.

\section{\label{sec:model}Models}
On the level of the $N$-particle quantum mechanics, the system can be described by the Lieb-Liniger(-McGuire)~\cite{LiebLiniger63,McGuire64} Hamiltonian with additional external potential $V_{\rm ext}$:
\begin{align}
\label{eq:H}
\hat{H} = &-\sum_{j=1}^{N}\frac{\hbar^2}{2m}\partial_{x_j}^2 + \sum_{j=1}^{N-1}\sum_{n=j+1}^{N}g_{1\rm D}\delta\left(x_j-x_{n}\right) \nonumber\\
&+\sum_{j=1}^{N}V_{\rm ext}\left(x_j\right)\;,
\end{align}
with $g_{1\rm D} < 0$.

 The mean-field equation corresponding to the Hamiltonian~(\ref{eq:H}) is the GPE~\cite{PethickSmith08}
\begin{align}
\label{eq:GPE}
i\hbar {\partial}_t\varphi(x, t) = &-\frac{\hbar^2}{2m}\partial_x^2 \varphi(x, t) +V_{\rm ext}(x) \varphi(x, t)\nonumber \\ &+(N-1)g_{1 \rm D}|\varphi(x, t)|^2 \varphi(x, t)\;,
\end{align}
where $|\varphi(x,t)|^2$ is the single-particle density (cf.~\cite{LiebEtAl00}).
{While the GPE~(\ref{eq:GPE}) often is used to describe Bose-Einstein condensates (BECs) of finite particle numbers, it can strictly speaking only be valid in the limit~\cite{LiebEtAl00}}: 
\begin{align}
\label{eq:limit}
\begin{array}{lcl}
N&\to&\infty\\ 
g_{\rm 1D}&\to&0
\end{array}
\quad  {\rm with}\quad Ng_{\rm 1D}&=\rm const.
\end{align}
 There even are cases for which the Gross-Pitaevskii functional becomes exact in this limit~\cite{LiebEtAl00}.

For attractive interactions and without an external potential [$V_{\rm ext}(x)\equiv 0$],
Eq.~(\ref{eq:GPE}) has exact soliton solutions of the form~\cite{PethickSmith08}
\begin{equation}
\label{eq:GPEsoliton}
\varphi(x,0) =\sqrt{\frac{2\mu}{(N-1)g_{\rm 1D}}}\frac{e^{im\mu x/\hbar-i(\mu-mu^2/2) t/\hbar}}{\cosh\left[\sqrt{\frac{2m|\mu|}{\hbar^2}}({x-x_0 -ut})\right]},
\end{equation}
where $u$ is the  velocity and $x_0$ the initial position. On the one hand, Eq.~(\ref{eq:GPEsoliton}) describes a single soliton, for which normalizing Eq.~(\ref{eq:GPEsoliton}) to one yields (cf.~\cite{CastinHerzog01})
\begin{equation}
\label{eq:GPE1soliton}
\mu = -\frac 18\frac{mg_{\rm 1D}^2}{\hbar^2}(N-1)^2\;.
\end{equation}
On the other hand, Eq.~(\ref{eq:GPEsoliton}) can also describe (well separated) parts of a solution. The sum of two such solutions which are, e.g., on both sides of a scattering potential corresponds to a fraction of the atoms being on one side and the rest of the atoms on the other side (note that the widths of the two solitons then depends on which fraction of the atoms is on each side). This is, however, very different from a quantum superposition of one soliton containing all atoms being simultaneously on both sides (cf.~\cite{WeissCastin09}).

Contrary to the localized mean-field solution~(\ref{eq:GPEsoliton}), eigensolutions of the $N$-particle Schr\"odinger equation corresponding to the Hamiltonian~(\ref{eq:H}) with zero external potential $V_{\rm ext}$ have to be translationally invariant (up to a phase factor):
\begin{equation}
\psi_N(\underline{x},t) \propto \exp\left(-\beta\sum_{j=1}^{N-1}\sum_{n=j+1}^{N}\left|x_j-x_{n}\right|+ik\sum_{j=1}^Nx_j\right)
\label{eq:Nsoliton}
\end{equation}
with 
\begin{equation}
\beta\equiv \frac{m|g_{1\rm D}|}{2\hbar^2}\;.
\end{equation}
The eigenenergy $E=E_0(N)+E_{\rm kin}$ of the eigenfunction~(\ref{eq:Nsoliton}) consists of the ground-state energy
\begin{equation}
\label{eq:E0}
E_0(N) =-\frac1{24}\frac{mg_{1\rm D}^2}{\hbar^2}N(N^2-1)
\end{equation}
and the
 center-of-mass kinetic energy
\begin{equation}
E_{\rm kin} = N\frac {\hbar^2 k^2}{2m} \;.
\end{equation}

On the one hand, there are cases for which mean-field and $N$-particle physics can be shown to agree:
By assuming the center-of-mass part of the wave function to be a delta function,
\begin{equation}
\psi_{\rm C}(X) = \delta(X-x_0-ut)\nonumber\;,
\end{equation}
 it is possible to show that the single-particle density of the Gross-Pitaevskii soliton~(\ref{eq:GPEsoliton}) and the many-particle soliton~(\ref{eq:Nsoliton}) coincide in the limit $N\gg 1$~\cite{CalogeroDegasperis75}. Thus, rather than simply being the solution of an approximated equation, the single-particle density of exact many-particle solitons~(\ref{eq:Nsoliton}) is well described by the mean-field case~(\ref{eq:GPEsoliton}).

On the other hand, there are cases for which there is no equivalent on the mean-field {(GPE)} level: Mesoscopic quantum superpositions.

\section{\label{sec:mesos}Mesoscopic quantum superpositions (MQS) vs.\ Hartree-product states}

As a starting point to investigate  mesoscopic quantum superpositions (MQS), this section first investigates parameter regimes accessible to the $N$-particle physics when scattering a bright quantum soliton off a potential by looking at the energetically allowed final states in the $N$-particle case (Sec.~\ref{sub:np}) after the soliton has left the potential. Section~\ref{sub:HP} investigates Hartree-product states of the form
\begin{equation}
\label{eq:HP}
\psi(\underline{x},t) = \prod_{j=1}^N \varphi(x_j,t)\;.
\end{equation}

\subsection{$N$-particle quantum mechanics\label{sub:np}}
It has been suggested that scattering slow bright quantum solitons of 
$N\approx 100$  particles off a scattering potential generates mesoscopic quantum superpositions~\cite{WeissCastin09,Streltsov09b}.  Scattering would produce a quantum superposition of all particles being either on one side of the scattering potential or at the other side:
\begin{equation}
\label{eq:NOON}
|\psi\rangle_{\rm MQS} = \frac 1{\sqrt{2}}\left(|N,0\rangle+e^{i\alpha}|0,N\rangle\right)
\end{equation} 
where the Fock-state notation $|N-n,n\rangle$ denotes $N-n$ particles on the right and $n$ particles on the left of the scattering potential. Such states have been called ``Schr\"odinger-cat states'' or ``NOON states''; further suggestions how mesoscopic quantum superpositions might be obtained can be found, e.g., in Refs.~\cite{CastinDalibard97, RuostekoskiEtAl98, DunninghamBurnett01,  MicheliEtAl03,  MahmudEtAl05,  Dounas-frazerEtAl07,  DagninoEtAl09, GertjerenkenEtAl10, GarciaMarchEtAl11, MazzarellaEtAl11,DellAnna12}.

MQS can be more general than Eq.~(\ref{eq:NOON}); quantum superposition involving states like
\begin{equation}
|\psi\rangle_{n} = \frac 1{\sqrt{2}}\left(|N-n,n\rangle+e^{i\alpha}|n,N-n\rangle\right), \quad n \lessapprox \frac N4
\end{equation}
or superpositions thereof are also interesting. {Furthermore, quantum superpositions which do not correspond to exact 50:50 splitting are also in the focus of research (cf.\ \cite{Streltsov09b}).}

\begin{figure}
\includegraphics[width=\linewidth]{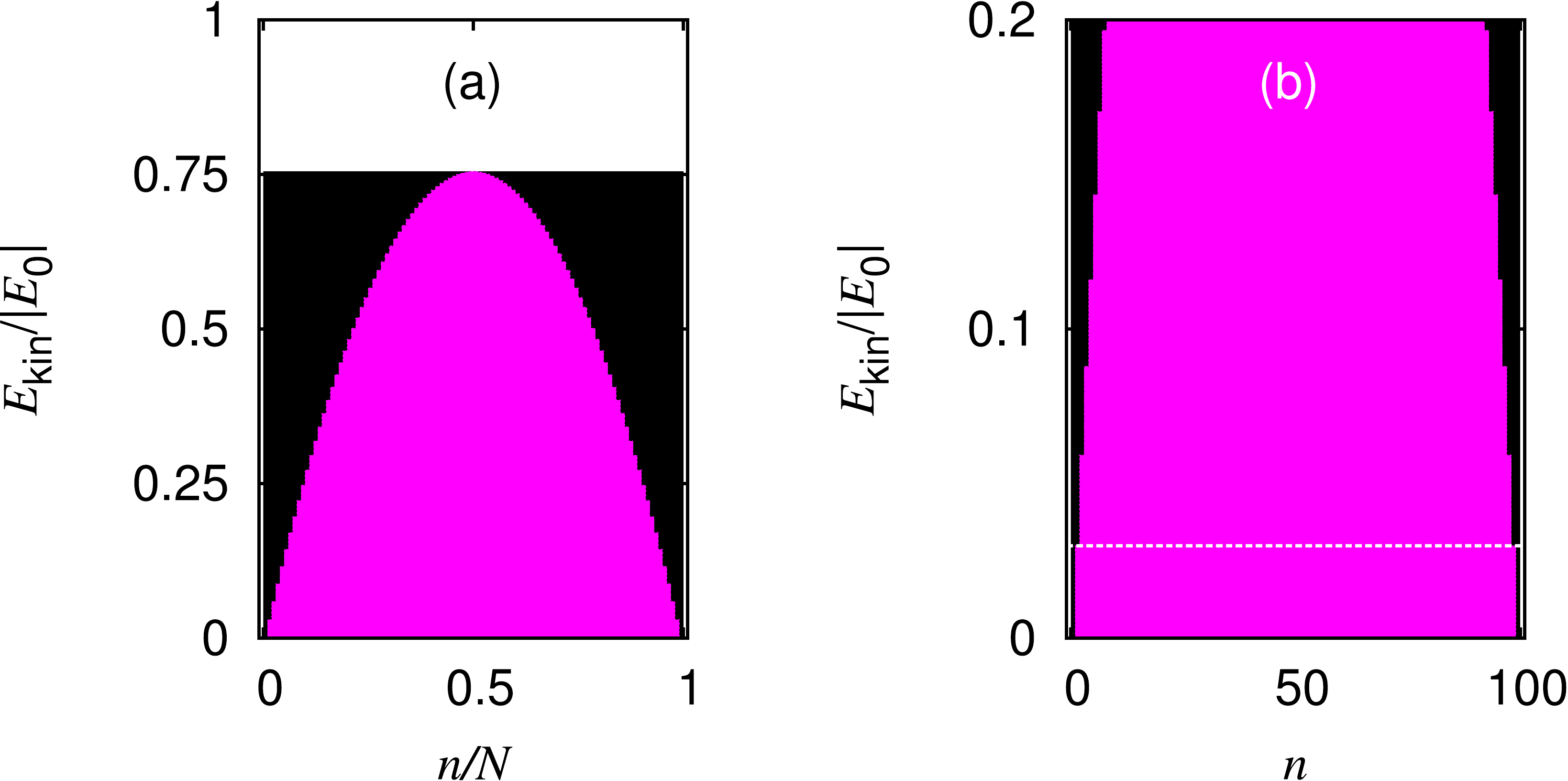}
\caption{\label{fig:phasedia}(Color online) Energetically (dis)allowed states for $N=100$ particles with  $n$ particles on one side of the scattering potential and  $N-n$ particles on other side. The parameter regions are displayed as a function of $n$ and center-of-mass kinetic energy $E_{\rm kin}$. \textbf{(a)} Magenta/gray region: not energetically allowed as soon as all particles have left the scattering potential, black and white regions: energetically allowed. Only in the white region [which lies above $E_{\rm kin}\simeq 0.755|E_0|$ for $N=100$; Eq.~(\ref{eq:boundary})] are product states with 50:50 occupation on both side of the barrier energetically allowed. \textbf{(b)} Enhanced lower part of the left panel. In the energy regime below the dashed line, the only energetically allowed states are: all particles on the left or all particles on the right of the barrier -- and quantum superpositions of both.}
\end{figure}

Figure~\ref{fig:phasedia} depicts the possible energy regimes. For high center-of-mass kinetic energy [well in the white area of Fig.~\ref{fig:phasedia}~(a), the parameter regime investigated, e.g.,  in Refs.~\cite{MartinRuostekoski2012,HelmEtAl2012}],
\begin{equation}
\label{eq:gg}
E_{\rm kin} \gg 2E_0\left(\frac N2\right)-E_0(N)\;,
\end{equation}
the soliton is energetically allowed to break into (at least) 2 parts.
The scattering potential can act as a beam splitter on the level of single particles. {In this energy regime, the mean-field soliton splits~\cite{MartinRuostekoski2012,HelmEtAl2012}, which includes 50:50 splitting. In the high energy regime, the particle radiation observable at lower energies disappears~\cite{HolmerEtAl07} and the system thus stays condensed.} Replacing the ``$\gg$'' by ``$=$'' in Eq.~(\ref{eq:gg}) yields the value for the boundary between the white and the black region in Fig.~\ref{fig:phasedia}~(a): 
\begin{equation}
\label{eq:boundary}
E_{\rm kin} \left\{
\begin{array}{lcl}
\simeq -0.755E_0 &:& N=100\\
= -0.75E_0 &:& N\to\infty
\end{array}\right.\;.
\end{equation}

For low center-of-mass kinetic energies [below the dashed line in Fig.~\ref{fig:phasedia}~(b)], 
\begin{equation}
\label{eq:perfect}
E_{\rm kin} < E_0(N-1) - E_0(N)\;,
\end{equation}
the soliton  is energetically forbidden to break into two parts and thus all particles are either on one side of the scattering potential ($n=0$) or on the other ($n=N$); quantum superpositions of both~(\ref{eq:NOON}) have been predicted~\cite{WeissCastin09}. 

In between the threshold given by Eqs.~(\ref{eq:boundary}) and (\ref{eq:perfect}), contributions of more states are allowed; the magenta/gray region indicates which $|N-n,n\rangle$ are energetically disallowed for each value of the center-of-mass kinetic energy. Contributions of states like $|N/2,N/2\rangle$ automatically imply that this part of the wave function has a higher energy than energetically allowed for any parameter regime which is labeled magenta/gray in Fig.~\ref{fig:phasedia}. This statement is independent of the $N$-particle state these particles are in as long as they cannot be found at the potential.

On the $N$-particle level governed by the Hamiltonian~(\ref{eq:H}), energy is conserved not only in the sense that $\langle\psi(t)| \hat{H}|\psi(t)\rangle$ is conserved but also in the sense that all higher moments $\langle\psi(t)| \hat{H}^{\nu}|\psi(t)\rangle$, $\nu=2,3,\ldots$ are time-independent\footnote{\label{page:footnote}As the time-evolution operator $U(t,0)=\exp(-i\hat{H}t/\hbar)$ (with $|\psi(t)\rangle = U(t,0)|\psi(0)\rangle$) commutes with $\hat{H}^{\nu}$ ($\nu=1,2,3,\ldots$), we have: $\langle\psi(t)| \hat{H}^{\nu}|\psi(t)\rangle =\langle\psi(0)| \hat{H}^{\nu}|\psi(0)\rangle $ for  $\nu=2,3,4,\ldots$. This excludes contributions from eigenfunctions for which the eigenenergy is not negative enough to contribute to the final state. Alternatively, $\hat{H}^{\nu}$ could be replaced by $[\hat{H}-E_0(N)]^{\nu}$ before repeating this analysis.}.
 Thus, within the magenta/gray parameter regime depicted in Fig.~\ref{fig:phasedia} energy conservation not only prevents the final state from being identical to $|N/2,N/2\rangle$, but $|N/2,N/2\rangle$ (and, $|N/2-n,N/2+n\rangle$ with increasing $n$ for decreasing initial center-of-mass kinetic energy) cannot even be an important contribution to the final wave function.

Just because a certain value for $n$ is energetically allowed does not necessarily imply that it will occur: at the threshold given by Eq.~(\ref{eq:boundary}), $n=N/2$ could, in principle occur. However, this would imply that in the final state, the fractions of the soliton on both sides of the barrier no longer move. It is thus more likely to occur for even higher initial center-of-mass kinetic energies.

\subsection{\label{sub:HP}Mean-field approach via Hartree-product states}
Often, mean-field theories are introduced~\cite{PethickSmith08} for bosonic $N$-particle quantum systems by starting with Hartree-product states~(\ref{eq:HP}),
which are subsequently used to derive Gross-Pitaevskii equations like Eq.~(\ref{eq:GPE})~\cite{PethickSmith08}. {This does, however, not necessarily imply that GPE is equivalent to the Hartree-product states; rather than interpreting the GPE-solution $\phi(x,t)$ automatically as being part of a Hartree-wave function, a more general approach is to {regard $|\phi(x,t)|^2$ as} the ``single-particle density''~\cite{LiebEtAl00}.}

Hartree-product states for which $\varphi(x,t)$ is zero on one side of the potential always exist. More interesting to compare with Sec.~\ref{sub:np} are wave functions for which $\varphi(x,t)$ is non-zero on both sides of the scattering potential. In this case, both the Fock-state $|N/2,N/2\rangle$ and $|N/2+n,N/2-n\rangle$ (with small $n$) are thus involved in the many-particle Hartree-product wave function: All Fock states contribute to the wave-function [as long as the single-particle wave function is non-zero at both sides, cf.~Eq.~(\ref{eq:choose})]. Note that contrary to the case of Hartree-product states, contributions of such states to the total wave function can be avoided in the case of a MQS. Strictly speaking, for low center-of-mass kinetic energies, those states are not accessible on the $N$-particle level (as discussed in the previous section). This seems to agree with the rather stepwise behavior of scattering mean-field solitons reported in Refs.~\cite{LeeBrand06,Streltsov09b}. 
This paper investigates the transition from this stepwise behavior to the continuous case similar to single-particle physics in more detail. We will show that Fig.~\ref{fig:phasedia} provides the energy scale on which the non-splitting GPE-soliton becomes a splitting soliton, depending on its initial center-of-mass energy.
The technical details are presented in Appendix~\ref{app:bound}.

\section{\label{sec:scattering}Scattering dynamics in the presence of harmonic confinement}

A recent (quasi-) 1D experiment~\cite{Hulet10,*Hulet10b} so far combines scattering solitons off a potential in 1D with additional harmonic confinement. We model this situation in Sec.~\ref{sub:HO}. So far, it does not yet realize the regime of both low kinetic energies and temperatures necessary to produce the quantum superpositions suggested in Refs.~\cite{WeissCastin09,Streltsov09b}.

For the $N$-particle quantum case (Sec.~\ref{sub:NPqp}),
the idea is to start with the many-particle ground state in the harmonic trap (cf.~\cite{HoldawayEtAl2012}),
the center of the trap can then (quasi-)instantaneously be shifted, followed by switching on the scattering potential in the middle of the trap. For the mean-field description via the GPE (Sec.~\ref{sub:GPE}) the same situation is repeated for a single soliton without initial center-of-mass kinetic energy. While the two Secs.~\ref{sub:NPqp} and \ref{sub:GPE} cover different energy regimes, Sec.~\ref{sub:beyond} demonstrates what effects can happen, in principle, on the $N$-particle level for the parameter-regime for which the GPE displays the peculiar behavior discussed in Sec.~\ref{sub:GPE}.

\subsection{\label{sub:HO}Harmonic confinement and scattering potential}
We assume the scattering potential to be narrow enough\footnote{Wide potentials were discussed in the $N$-particle quantum case in Refs.~\cite{WeissCastin09,Streltsov09b}.} for it to be approximated by a delta function:
\begin{equation}
\label{eq:Vext}
V_{\rm ext}(x)= \frac12 m \omega^2 x^2 + v_0\delta(x)\;;
\end{equation}
where ``narrow'' refers to the potential being narrower than both the soliton width~(cf.~\cite{KivsharEtAl90,Muller11}) and the oscillator length:
\begin{equation}
\label{eq:lambdaGPE}
\lambda_{\rm GPE} \equiv \left(\frac{\hbar}{m\omega}\right)^{\frac 12}\;, 
\end{equation}
where the index GPE indicates that this is a relevant length scale for the GPE.

\subsection{\label{sub:NPqp}$N$-particle quantum physics: effective potential approach}

The effective potential approach of Ref.~{\cite{WeissCastin09,SachaEtAl09}}  is valid for low center-of-mass kinetic energies;\footnote{{The validity of the effective-potential approach~\cite{WeissCastin09,SachaEtAl09} {to describe time-dependent scattering} has been proved rigorously by calculating strict bounds on the transmission and reflection amplitudes~\cite{WeissCastin12}.}} the system can be described by an effective Schr\"odinger equation for the center-of-mass coordinate~$X$. This effective Schr\"odinger equation reads:
\begin{align}
\label{eq:schrodingerCoM}
i\hbar \partial_t\psi_{\rm C}(X,t) =& \left[-\frac {\hbar^2}{2Nm}\partial_X^2 + \frac12Nm\omega^2X^2\right]\psi_{\rm C}(X,t)\nonumber\\
&+V_{\rm eff}(X)\psi_{\rm C}(X,t)
\end{align}
where the effective potential is given by~\cite{WeissCastin09,SachaEtAl09}
\begin{align}
V_{\rm eff}(X) &=\int d^Nx{|{\psi}_{N,k}(\underline{x})|^2}{V(\underline{x})}\delta\left({\textstyle X\!-\!\frac1N\sum_{\nu=1}^Nx_{\nu}}\right)\nonumber\\
&\equiv\frac{U_0}{\cosh^2(X/\ell)}\;,
\label{eq:Veff}
\end{align}
where the last line uses the results of Ref.~\cite{CalogeroDegasperis75} and the parameters are:
the strength
\begin{equation}
U_0\equiv \frac {Nv_0}4\frac{m|g_{\rm 1D}|(N-1)}{\hbar^2},
\end{equation}
which is the product of the particle number, the single-particle potential strength $v_0$ [Eq.~(\ref{eq:Vext})] and the soliton amplitude [Eqs.~(\ref{eq:GPEsoliton}) and (\ref{eq:GPE1soliton})] and the
width
\begin{equation}
\ell \equiv 2\frac{\hbar^2}{m|g_{\rm 1D}|(N-1)}
\end{equation}
 given by the soliton width~\cite{CastinHerzog01}, cf.\ Eq.~(\ref{eq:GPE1soliton}).

The effective potential~(\ref{eq:Veff}) thus has the form of the soliton; the ratio of the width~$\ell$ to the center-of-mass oscillator length [cf.~Eqs.~(\ref{eq:lambdaGPE}) and (\ref{eq:schrodingerCoM})]
\begin{align}
\lambda_{\rm  osc} &\equiv \left(\frac{\hbar}{Nm\omega}\right)^{\frac 12}  \\
&= \frac{\lambda_{\rm GPE}}{\sqrt{N}},
\end{align}
influences the physics.

As soon as the center-of-mass density near the effective potential approaches zero, MQS states emerge (cf.~\cite{WeissCastin09}). For center-of-mass energies which are an order of magnitude higher, numerics still predicts quantum superpositions far from product states~\cite{Streltsov09b}.

One advantage of the additional harmonic trapping potential is that excitations due to an opening of the trap which prepares the initial state~\cite{Castin09} can be discarded. Another advantage is discussed in this section: simply producing potentially interesting MQS and showing that in an experiment all particles are either on one side of the potential or at the other would not be enough to confirm the existence of MQS; additional experimental signatures are necessary. One possibility is to look at the interference in the center-of-mass density after removing the barrier~\cite{WeissCastin09}.

\begin{figure}
\includegraphics[width=\linewidth]{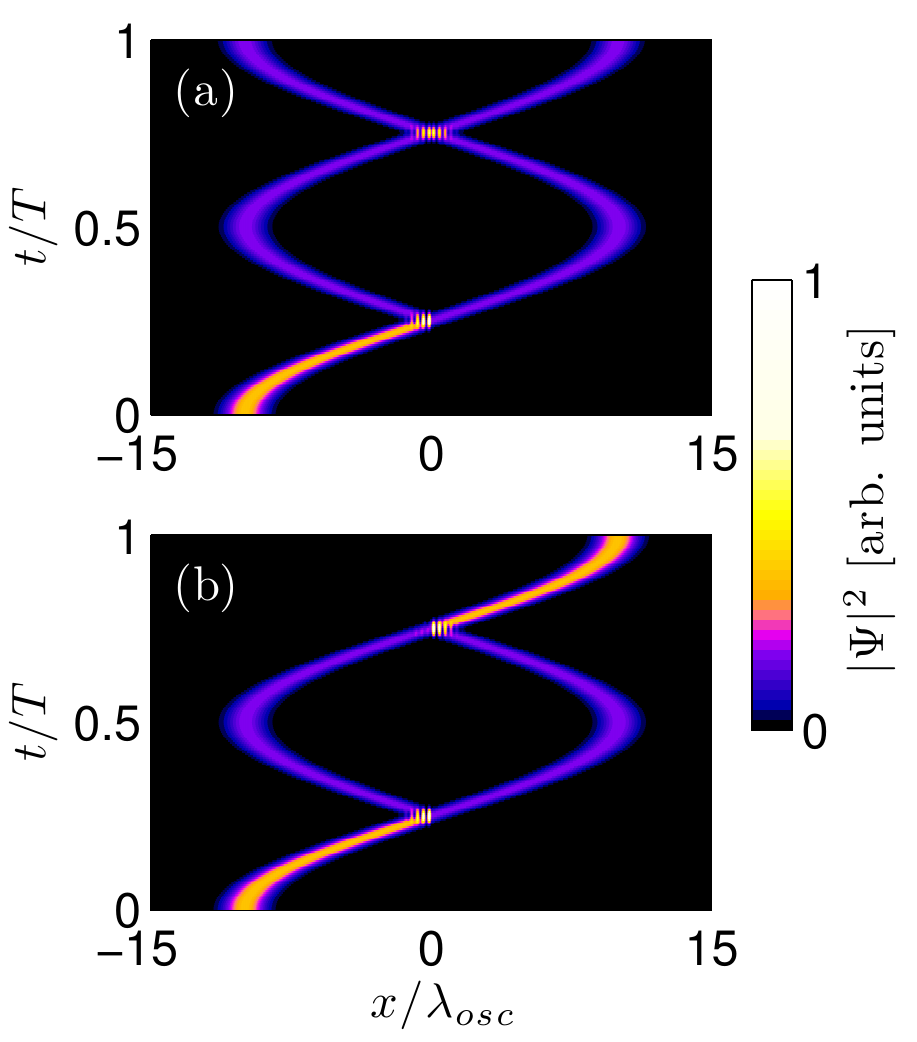}
\caption{\label{fig:signature}(Color online) Center-of-mass density of a 1D bright quantum-matter-wave soliton  in a two-dimensional projection as a function of both time $t$ (in units of the oscillation period $T$) and elongation $x$ (in units of the oscillator length $\lambda_{\rm osc}$). There is additional harmonic confinement ($\propto x^2$); the soliton is modeled within the effective potential approach.  Initially, the wave function is centered around $x=-10\lambda_{\rm osc}$. At $t = 0.25T$ the soliton scatters off a narrow effective potential modeled by a delta potential and the center-of-mass wave function splits into two parts, leading to a mesoscopic quantum superposition of all particles either being on the left or right of the scattering potential at $t=0.5T$. \textbf{(a)} If the quantum superposition becomes a statistical mixture due to decoherence at $t\approx 0.5T$, there is a $50\%$ probability to find the particles on either side of the scattering potential at $t=T$. \textbf{(b)} For a quantum superposition, all particles end on the right side with $p_{\rm r}=98.5\%$~probability at $t=T$.}
\end{figure}

For not-too-broad effective potentials~(\ref{eq:Veff}), there is a simpler approach shown in Fig.~\ref{fig:signature}:
After the second collision of both parts of the wave function in the presence of the barrier, this leads to a probability close to one of all particles being  at the side opposite to the initial condition [Fig.~\ref{fig:signature}~(b)], which considerably differs from the case of a statistical mixture [Fig.~\ref{fig:signature}~(a)]. Losing a single particle after the creation of the quantum superposition would turn the MQS~(\ref{eq:NOON}) into such a mixture. {While the precise influence of decoherence will depend on experimental details, we model decoherence by replacing the quantum superposition of all atoms being either on one side of the scattering potential or on the other by a statistical mixture near $t\approx T/2$}.

\begin{figure}
\includegraphics[width=\linewidth]{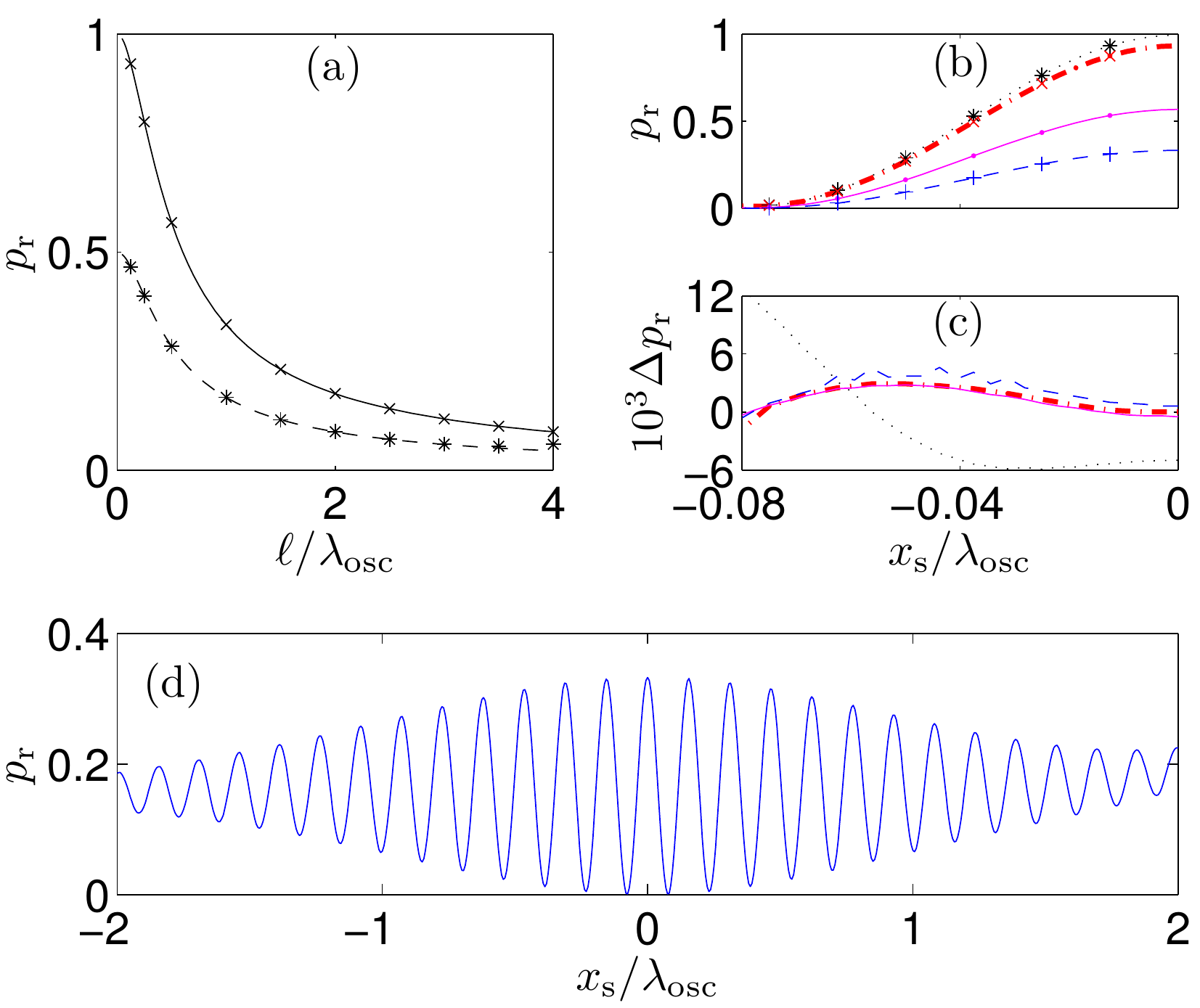}
\caption{\label{fig:interfero}(Color online) \textbf{(a)} For the same situation as in Fig.~\ref{fig:signature}, the probability to find all particles on the right side is plotted as a function of the width of the effective potential~(\ref{eq:Veff}). The cases of the quantum superposition (solid line) and the statistical mixture (dashed line) are clearly distinguishable for repeated measurements. Points: Averaging over an experimentally realistic~\cite{Gross11} narrow Gaussian distribution with width~$5$ centered around $N=100$ and truncated at 90 and 110. \textbf{(b)} Probability $p_r$ to find  all particles on the right of the scattering potential as a function of shift $x_s$. From top to bottom: $\ell/\lambda_{\rm osc}=0, 0.125, 0.5, 1.0$. \textbf{(c)} Difference of numeric solution and the computer-algebra based analytic solution extending~(\ref{eq:T2}) to general potentials~(\ref{eq:Veff}) for the same parameters as in the previous panel.  \textbf{(d)} Probability to find all particles on the right hand side as a function of the shift for $\ell=\lambda_{\rm osc}$. Compared to Eq.~(\ref{eq:T2}), the amplitude is lower than one (caused by the wider potential). Furthermore, the fact that the shift of the potential leads to an interference of only parts of the wave packets [discussed below Eq.~(\ref{eq:T2})] is clearly visible.}
\end{figure}

In order to understand this behavior, let us first assume the scattering potential is narrow enough to be approximated by a delta function:
\begin{equation}
V_{\rm eff}(X) = \frac{\hbar^2}{m}\Omega\delta(x+x_s).
\end{equation}
where we included a small shift to the left.

In order to derive the leading order behavior of our center-of-mass wave packet, start with the fact that without the scattering potential, after some time the wave packet would be in the middle of the potential with all the initial potential energy being transformed into kinetic energy. For a plane wave with wavevector $k$, the transmission coefficient is given by~\cite{Fluegge90}
\begin{equation}
T =
\frac{ik}{ik-\Omega}
\end{equation}
 and reflection coefficient by~\cite{Fluegge90}
\begin{equation}
R =
\frac{\Omega}{ik-\Omega}\;.
\end{equation}
For 50:50 splitting, they must have the same modulus, which thus defines the strength of the potential:
\begin{equation}
\label{eq:onehalf}
\Omega \equiv k\;.
\end{equation}
In our case, we do not have plane waves but wave packets centered around $k=\Omega$. The time scales for the reflection of both parts of the wave packet in the harmonic potential are the same. 

Due to the harmonic confinement, in the further time evolution a second scattering takes place where it has to be considered that the transmitted part had to cover an additional distance of $4x_s$:
\begin{align} \label{eq:wavefunc}
u(x)=\begin{cases}
  \frac{{e}^{{i}\Omega x}}{{i}-1} + \frac{{e}^{-{i}\Omega x}}{\left({i}-1\right)^2} +\frac{{i}^2{e}^{-{i}\Omega x} {e}^{-4{i}\Omega x_s}}{\left( {i}-1\right)^2} &:\; x<-x_s\\
  \frac{{i}{e}^{{i}\Omega x}}{\left({i}-1\right)^2}+ \frac{{i}{e}^{-{i}\Omega x-4{i}\Omega x_s}}{{i}-1} + \frac{{i} {e}^{{i}\Omega x-4{i}\Omega x_s}}{\left({i}-1 \right)^2} & :\; x>-x_s.
\end{cases}
\end{align}
The transmission coefficient after two reflections is therefore given by
\begin{align}
\label{eq:T2}
T & =  \left|\frac{{i}}{\left({i}-1\right)^2} + \frac{{i}}{\left({i}-1 \right)^2}{e}^{4{i}\Omega x_s} \right|^2\nonumber\\& = \frac{1}{2}\left[1+\cos\left(4\Omega x_s \right)\right]\;.
\end{align}
Thus, if the scattering potential can be approximated by a delta function, we can indeed expect a probability close to one that all particles are on the right side of the scattering potential if they initially were on the other side. A more complete analysis would have to include wave packets rather than plane waves. This will effectively damp the oscillation amplitude as a function of the displacement $x_s$ as can be seen in Fig.~\ref{fig:interfero}~(d). In this panel, the probability to find all particles on the right does not reach 1 because a broader effective potential (cf.~\cite{WeissCastin09}) was used.

In Fig.~\ref{fig:interfero} we display the probability to find all particles on one side of the potential after one oscillator period. The quantum mechanics of pure states can clearly be
distinguished from statistical mixtures over a wide parameter regime.
The fact that the above scheme depends on the position of the scattering potential offers a 
potential application to interferometry by identifying small potential gradients along the center of the harmonic trap  (for applications of ultra-cold atoms in interferometric experiments discussed in the literature see, e.g., \cite{DimopoulosEtAl07,GrossEtAl10,MartinRuostekoski2012,HelmEtAl2012}).

 While the difference between pure quantum dynamics and statistical mixture is particularly large for a very narrow scattering potential, it is still clearly visible for broader effective potentials [Fig.~\ref{fig:interfero}]. The two values are distinguishable as soon as the error of the means are small enough (which scale as $1/\sqrt{N_{\rm rep}}$, where $N_{\rm rep}$ is the number of repetitions of the experiment). The experimentally realizable parameters for a soliton of $N=100$ particles discussed in Ref.~\cite{WeissCastin09} correspond to $\ell=1.5\lambda_{\rm osc}$ -- where the difference between statistical mixture and MQS is still clearly visible.

\subsection{\label{sub:GPE}Mean-field approach via the GPE}

Modeling the same situation as in Sec.~\ref{sub:NPqp} on the level of the GPE leads to a different energy regime because of the mean-field limit~(\ref{eq:limit}). In this limit, the effective potential regime below the dashed line in Fig.~\ref{fig:phasedia}~(b) could only be covered for vanishing ratio of center-of-mass kinetic energy to ground-state energy. However, the GPE can cover the regime for which this ratio is finite.

Based on the reasoning of Sec.~\ref{sub:HP} (cf.~Sec.~\ref{sub:diff}), it would not be surprising to find a stepwise behavior of the reflection (or transmission) coefficient which jumps from 0 to 1.

\begin{figure}
\includegraphics[width=\linewidth]{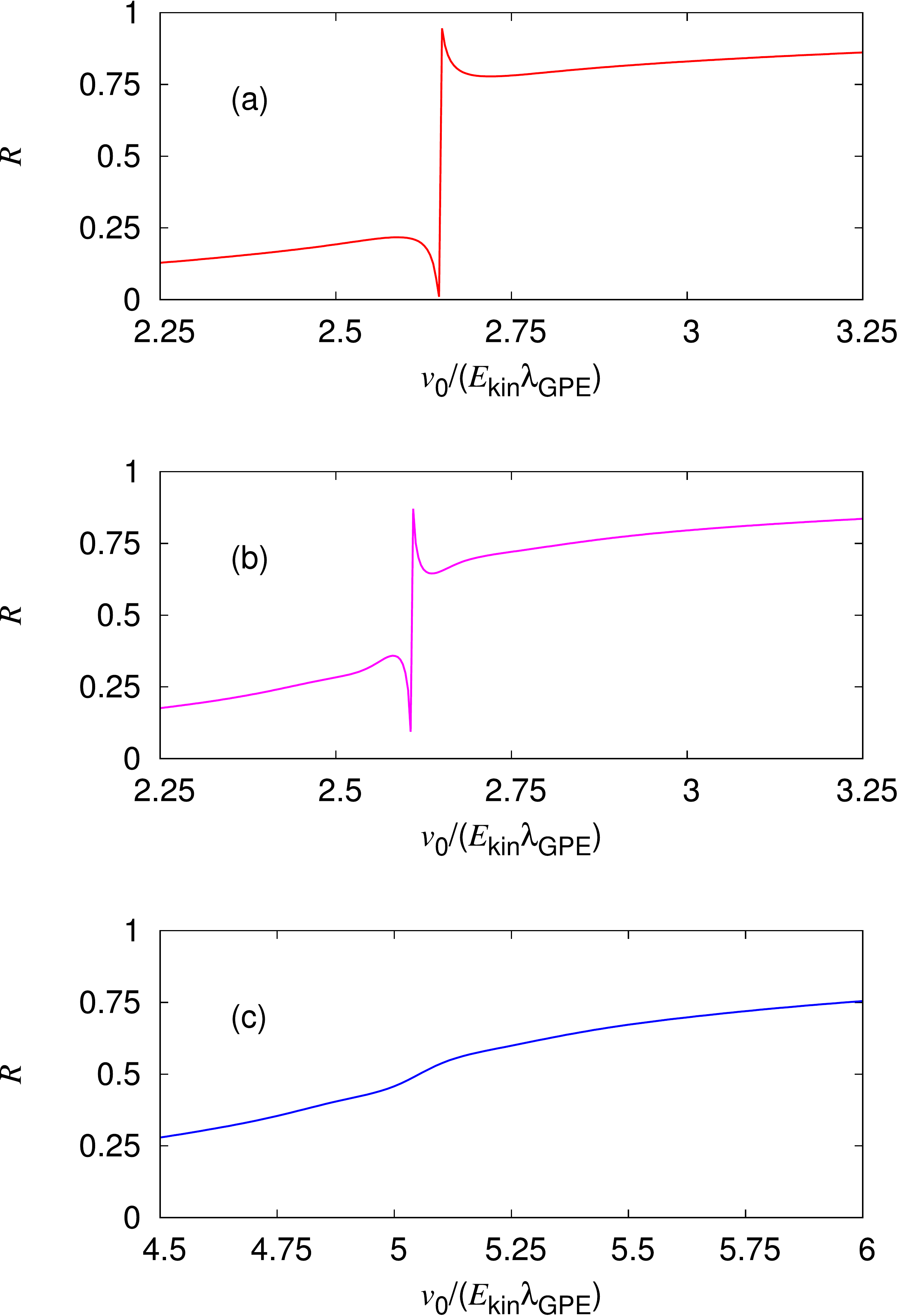}
\caption{\label{fig:GPHO}(Color online) Reflection of a GPE-soliton in a 1D harmonic trap from a narrow Gaussian barrier [used to model the delta-potential in Eq.~(\ref{eq:Veff})] as a function of strength of the scattering potential. The initial center-of-mass kinetic energy increases from top to bottom. (a) $E_{\rm kin}= 0.7|E_0|$: The reflection $R$ remains below 0.22 before jumping to values above 0.77 near $v_0/(E_{\rm kin}\lambda_{\rm GPE})=2.65126$. (b) $E_{\rm kin}= 0.8|E_0|$: The reflection $R$ remains below 0.36 before jumping to values above 0.64. (c) $E_{\rm kin}= 1.0|E_0|$: The reflection $R$ is continuous.}
\end{figure}

Figure~\ref{fig:GPHO} shows a different behavior: while there are indeed jumps in the reflection coefficient if the strength of the scattering potential is increased, this jump lies below 1 for many parameters and gets smaller if the kinetic energy approaches the threshold~(\ref{eq:boundary}). When repeating the calculation without harmonic confinement, the qualitative behavior in all three cases is the same.

Furthermore, changing from energies for which the product state corresponding to 50:50-splitting   cannot exist on both sides of the scattering potential [Fig.~\ref{fig:GPHO}~(a)] to energies for which it can (just) exist [Fig.~\ref{fig:GPHO}~(b)] primarily reduces the size of the gap (for Fig.~\ref{fig:GPHO} by a factor of 2) before it eventually vanishes for even higher kinetic energies. In order to investigate this in more detail, the next section focuses on a more detailed analysis without the harmonic confinement.

\subsection{\label{sub:beyond}$N$-particle quantum physics: beyond the effective potential approach}
One approach to include more particles is to discretize the $N$-particle Schr\"odinger equation corresponding to the Hamiltonian~(\ref{eq:H}), {which is a delicate matter for attractive systems}. This leads to a Bose-Hubbard model,
\begin{align}
\label{eq:BH}
\hat{H}_{\rm discretized} = 
&-J \sum_{j}\left(\hat{c}_j^{\dag}\hat{c}_{j+1}^{\phantom{\dag}}
+\hat{c}_{j+1}^{{\dag}}\hat{c}_{j}^{\phantom{\dag}}\right)\nonumber\\
&+\frac U2 \sum_{j}\hat{n}_j\left(\hat{n}_j-1\right)
\nonumber\\
&+A  \sum_{j}\hat{n}_j j^2 + \widetilde{v}_0\delta_{j,0}
\;,
\end{align}
where $\hat{c}_j^{(\dag)}$ are the boson creation and annihilation operators
on site $j$, $\hat{n}_j = \hat{c}_j^{\dag}\hat{c}_j^{\phantom{\dag}}$ are the number operators, $J$ is the hopping matrix element and $U$ the on-site interaction energy. For a small Hilbert space, such a model can be solved via exact diagonalization (see, e.g, \cite{SorensenEtAl12} and references therein); for a larger Hilbert space, imaginary time evolution is a much better choice to determine the ground state, that is our initial condition (cf.\ \cite{GlickCarr11}). We use this to find our initial condition and subsequently numerically solve the { full time-dependent Schr\"odinger equation corresponding to the Hamiltonian~(\ref{eq:BH})}
 {via the Shampine-Gordon routine~\cite{ShampineGordon75}}.

While the limit~(\ref{eq:limit}) is not accessible on the $N$-particle level, $N=4$ still allows us to use the full Hamiltonian and get physical insight into what happens on the $N$-particle quantum level. For future quantitative comparison with experiments, advanced approximate methods on the $N$-particle level as used in Refs.~\cite{Streltsov09b,GlickCarr11} will be useful.

\begin{figure}
\includegraphics[width=\linewidth]{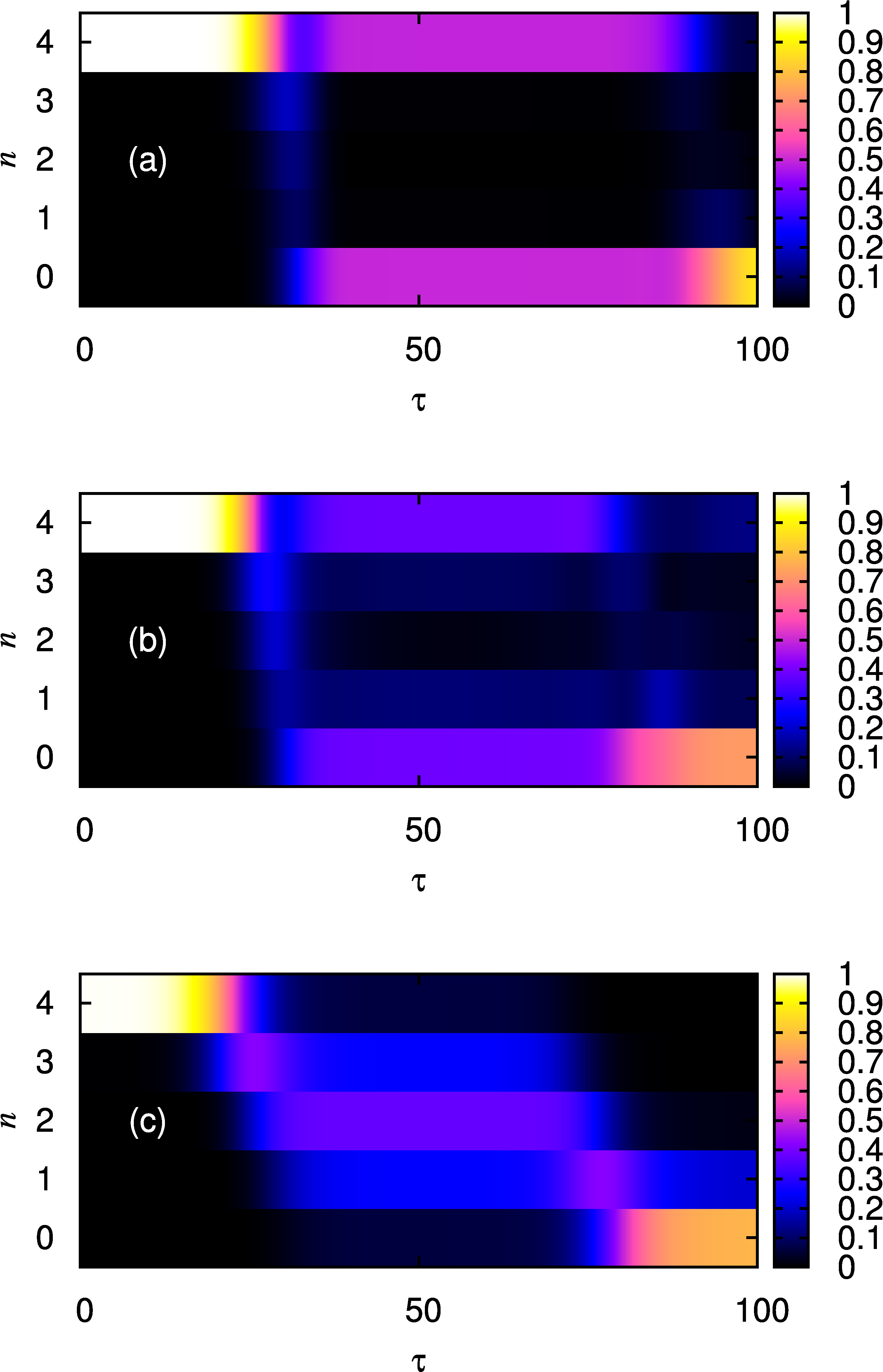}
\caption{\label{fig:BH}(Color online) Probability to find $n$ particles on the left side of the scattering potential as a function of time $\tau=Jt/\hbar$ for a total particle number of $N=4$. Initially, the system [modeled by Eq.~(\ref{eq:BH}) on 51 lattice sites] is in the $N$-particle ground state (determined via imaginary time-evolution) of a one-dimensional harmonic oscillator shifted by $12$ lattice sites before removing the shift and switching on the scattering potential. With strong attractive interactions [$NU/J=-4$, panel (a)], the probability to find 4 particles on either side of the scattering potential lies above 98\% for $\tau\approx 50$. For weaker interactions [$NU/J=-3$, panel (b)] this value has already dropped to less than $78\%$. Without interaction [$NU/J=0$, panel (c)], the value has dropped to $12.5\%$, the value predicted for product states [see Eq.~(\ref{eq:choose})]. The values for the scattering potential ($\widetilde{v}_0/J=\{0.27,0.43,0.82\}$) were chosen via bisection; $A=0.001J$.}
\end{figure}

Figure~\ref{fig:BH} shows the difference between the regime of perfect MQS like~(\ref{eq:NOON}) in the regime of low center-of-mass kinetic energy [Fig.~\ref{fig:BH}~(a), cf.\ Sec.~\ref{sub:NPqp}] compared with the high kinetic energy regime [Fig.~\ref{fig:BH}~(c)] and the regime of medium kinetic energies  [Fig.~\ref{fig:BH}~(b)]. In the high-energy regime, the final state corresponds to a product state for which the probability distribution of particle numbers always has a single peak.
Bimodal distributions correspond to quantum superpositions of (for small particle numbers) primarily two states,
$|n_1,N-n_1\rangle$ and $|n_2,N-n_2\rangle$  for which the $n_1$ and $n_2$ differ by more than~1.

\section{\label{sec:bound}Bounds and quantitative predictions for the GPE-solutions}
Without the harmonic confinement, the qualitative behavior is the same as in Sec.~\ref{sub:GPE}; in addition, as soon as the probability to find particles on the scattering potential is small, the exact eigensolutions of the Lieb-Liniger equations are sufficient to expand the wave function.

Including center-of-mass kinetic energy and using again the notation that $n$ particles are on one side and $N-n$ on the other side of the scattering potential, we can use energy conservation:\footnote{{We again assume that the wave function vanishes at the potential. This allows us to use the exact eigenfunctions of the Lieb-Linger model, cf.~Eq.~(120) of Ref.~\cite{CastinHerzog01}.}}
\begin{equation}
  E_{\rm kin} + E_0(N) = {E}_{{\rm kin}, 2} + E_0(n)+E_0(N-n).
\label{eq:conserve} 
\end{equation}
While the left-hand side describes the initial condition, the final situation does not necessarily have to consist of exactly two solitons as the soliton is energetically allowed to break into more than two parts. This does not influence the exact bounds we derive in this section but it might affect quantitative predictions derived from Eq.~(\ref{eq:conserve}). While this section does not explicitly involve Hartree-product states, Appendix~\ref{app:bound} shows that a Hartree product state leads to replacing the variable $n$ in Eq.~(\ref{eq:conserve}) by its mean $\overline{n}$. After this replacement, the derivations of this section remain valid even for Hartree-product states.

Dividing Eq.~(\ref{eq:conserve}) by $|E_0(N)|$, and taking the {mean-field limit~(\ref{eq:limit})} yields:
\begin{align}
\label{eq:Ekin2}
 \frac{{E}_{\rm kin}-{E}_{{\rm kin}, 2}}{|E_0(N)|}
&= +1-R^3-(1-R)^3\nonumber\\
&= 3R(1 - R)
\end{align}
where
\begin{equation}
\label{eq:RnN}
R = \frac nN
\end{equation}
is the fraction of the particles which are reflected. Note that this argument assumes that the soliton breaks into exactly two parts - for more than two parts the final center-of-mass kinetic energy ${E}_{{\rm kin}, 2}$ would be even lower.

As the final kinetic energy cannot be negative, Eq.~(\ref{eq:Ekin2}) implies
\begin{align}
\label{eq:EkinGreater}
 \frac{{E}_{\rm kin}}{|E_0(N)|}\ge 3(1-R)R\;.
\end{align}
Thus, for small kinetic energies not all values for $R$ are allowed. To proceed, we put all $R$ into the sets $\{R\le 0.5\}$ and  $\{R\ge 0.5\}$ (or in both in case the value $R=0.5$ occurs). Defining
\begin{equation}
\label{eq:Rmin}
R_{\rm min} = \min\{R\ge 0.5\}
\end{equation}
and
\begin{equation}
\label{eq:Rmax}
R_{\rm max} = \max\{R\le 0.5\}
\end{equation}
implies that for very low center-of-mass kinetic energy, $R\approx 1$ or $R\approx 0$ are possible -- contrary to the quantum case which (depending on the strength of the scattering potential) allows all values for $R$ in the range $0\le R\le 1$.

\begin{figure}
\includegraphics[width=\linewidth]{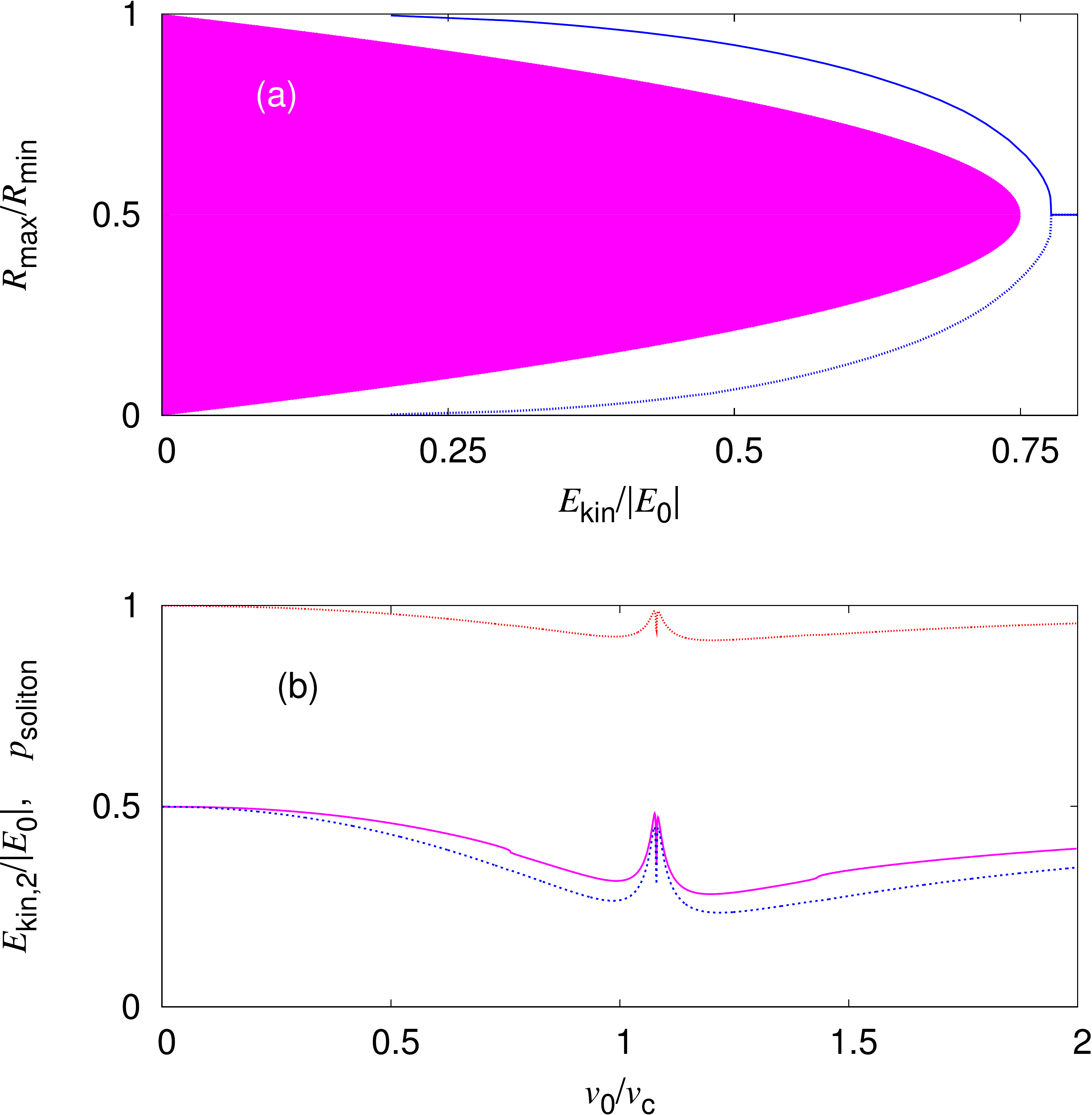}
\caption{\label{fig:GP}Scattering a mean-field soliton off a delta potential in a 1D situation without additional harmonic confinement. \textbf{(a)} The reflection $R_{\rm min}$ (solid line) and  $R_{\rm max}$ (dotted line) 
as defined in Eqs.~(\ref{eq:Rmin}) and (\ref{eq:Rmax}) as a function of initial center-of-mass kinetic energy. As predicted by Eq.~(\ref{eq:EkinGreater}), both curves lie outside the magenta/gray region for energetic reasons. \textbf{(b)} Scattering of a GPE soliton with initial center-of-mass kinetic energy $E_{\rm kin}=0.5|E_0|$ as a function of the strength of the delta-scattering potential, normalized such that without interaction, there would be 50:50 splitting [cf.\ Eq.~(\ref{eq:onehalf})]. From top to bottom: Dotted red/black line corresponds to the the fraction $p_{\rm soliton}$ of the final GPE-solution which is in solitons (the rest being particle-radiation). Solid magenta/gray line:  Total kinetic energy after the scattering as predicted by Eq.~(\ref{eq:Ekin2}) using the numerically calculated values for $R$. Dashed blue/black line: numerically calculated total kinetic energy. The difference between the two lower lines is only large if $p_{\rm soliton}<1$.}
\end{figure}

To test this statement, the magenta/gray region in Fig.~\ref{fig:GP}~(a) shows the energetically forbidden parameter regime for $R$ as described by Eq.~(\ref{eq:EkinGreater}). Numerically calculating the values for $R_{\rm max}$ and $R_{\rm min}$ yields that they indeed lie outside this regime. The jumps of the reflection reported in Refs.~\cite{LeeBrand06,Streltsov09b} are also visible in this figure; we can quantify that it takes place for center-of-mass kinetic energies below one fourth of the modulus of the ground-state energy.

In order to test the prediction of Eq.~(\ref{eq:conserve}), the center-of-mass kinetic energy of the solitons after scattering off the potential has to be evaluated numerically. To do this, we perform a numerical scattering transform as described in Ref.~\cite{boffetta_osborne_jcp_1992}. Figure~\ref{fig:GP}~(b) demonstrates that Eq.~(\ref{eq:conserve}) gives a good qualitative description of the final center-of-mass kinetic energy, but there is no perfect agreement. This can be explained by numerically calculating, again using a numerical scattering transform, the fraction of the final wave function which is contained in solitons, the rest being ``radiation'' of single particles having left the soliton. Panel (b) also displays an interesting behavior near the discontinuity of the reflection coefficient which manifests itself in a much better agreement of the two lower curves accompanied by nearly the whole GPE wave function being contained in solitons.

\section{\label{sec:concl}Conclusion}
We investigate scattering bright solitons generated from potentials both on the mean-field level and on the $N$-particle quantum level in 1D.  Adding an additional harmonic confinement leads to interesting effects both on the $N$-particle quantum level and on the GPE mean-field level: 

Firstly, adding a harmonic confinement to the creation of quantum superpositions of slow bright quantum matter-wave solitons provides a possibility to distinguish quantum superpositions from statistical mixtures: after scattering off the potential twice, the probability that all particles are on the side of the potential opposite to the initial condition clearly distinguishes the two cases. As on the single-particle level, changing the strength of the potential leads to a continuously varying reflection (or transmission) coefficient. 

Secondly, for the reflection behavior of Gross-Pitaevskii solitons, we have derived analytic bounds on the size of the jump of the reflection coefficient and derived the energy scale on which the step vanishes continuously for increasing center-of-mass kinetic energy. This bridges the two types of behavior previously reported in the literature: bright solitons on the Gross-Pitaevskii level have been reported to split for larger energies (cf.\ \cite{MartinRuostekoski2012,HelmEtAl2012}) and display a stepwise reflection-behavior for lower center-of-mass kinetic energies~\cite{LeeBrand06,Streltsov09b} for which the GPE-soliton hardly splits. 

 While the nonlinear character of the GPE does not allow GPE-solutions which are quantum superpositions, we conjecture that the jumps in the transmission behavior of the GPE-reflection coefficient indicates the formation of interesting quantum superpositions at the N-particle quantum level.

{
\paragraph*{Note added:} the jumps on the GPE-level were recently also investigated in Ref.~\cite{WangEtAl12}.}

\acknowledgments

We thank Y.~Castin, S.~A.~Gardiner, J.~L.~Helm,  D.~I.~H.~Holdaway, M.~Holthaus, C.~L\"ammerzahl and E.~Zaremba for discussions. C.W.\ thanks R.~G.\ Hulet for discussing the experiment~\cite{Hulet10,*Hulet10b}. We thank the \textit{Studienstiftung des deutschen Volkes} (B.G.) and the \textit{Heinz Neum\"uller Stiftung} (B.G.), 
 Durham
University (T.B.) and the UK EPSRC (Grant No. EP/G05
6781/1, T.B.\ and C.W.) for funding.
Computer power was obtained from the GOLEM and HERO  cluster
of the University of Oldenburg and the cluster Hamilton at Durham University. B.G.\ thanks C.\ S.\ Adams and S.\ A.\ Gardiner for the hospitality during her visits at Durham University. 

\begin{appendix}
\section{\label{app:bound} Re-deriving the bounds of Sec.~\ref{sec:bound} for Hartree-product states in the limit  $N\to\infty$}
This section deals with Hartree-product states~(\ref{eq:HP}) for which the  single-particle wave function~$\varphi(x,t^*)$ is zero near the scattering potential and non-zero on both sides of the scattering potential at a certain point of time~$t^*$. Using properties of the eigenfunctions of the Lieb-Liniger model~\cite{CastinHerzog01} (Sec.~\ref{sub:eigen}), we extend the calculations of Sec.~\ref{sec:bound} to Hartree-product states. While this could, in principle, lead to different results, in the limit $N\to\infty$ we recover the same bounds as before (Sec.~\ref{sub:appbound}). Only by assuming that the higher energy states not accessible on the $N$-particle quantum level are also inaccessible to Hartree-product states (cf.\ the footnote on page~\pageref{page:footnote}), Hartree-product states would predict a different reflection behavior than observed at the GPE-level (Sec.~\ref{sub:diff}).

\subsection{\label{sub:eigen}Justification why the eigenfunctions of the Lieb-Liniger Model can be used}
We take the  scattering potential to have the form
\begin{equation}
V_{\rm scatt}(x) = 0 \quad {\rm for}  \quad |x| > \zeta\;,
\end{equation}
which is fulfilled exactly for the delta function used in this paper and approximately for Gaussian or $1/\cosh^2(x)$ potentials.

The single-particle wave function defining the Hartree-product states~(\ref{eq:HP}) is assumed to fulfill:
\begin{equation}
\label{eq:null}
0 = \int_{-\zeta}^{\zeta}|\varphi(x,t^*)|^2 dx\;.
\end{equation}
This requirement was fulfilled in the numerics of Sec.~\ref{sec:bound}; in the following we use that
\begin{equation}
\label{eq:pdeff}
p \equiv \int_{\zeta}^{\infty}|\varphi(x,t^*)|^2 dx
\end{equation}
implies
\begin{equation}
\int_{-\infty}^{-\zeta}|\varphi(x,t^*)|^2 dx = 1-p\;.
\end{equation}

Equation~(\ref{eq:null}) implies that we can use the eigenfunctions of the Lieb-Liniger model to expand the wave-function. If $n$ particles are on one side, they can form an $n$-particle soliton with energy:
\begin{equation}
\langle E\rangle = \langle E_{\rm kin}\rangle  + E_0(n)
\end{equation}
where $E_0$ is given by Eq.~(\ref{eq:E0}). Alternatively, it could form several solitons or even free particles. However, any such combination has a higher energy $E_0$ as $E_0(n+\widetilde{n})\leq E_0(n)+E_0(\widetilde{n})$.

Thus, if $n$ particles are on one side of the scattering potential and $N-n$ on the other (which we denote by $| n, N-n\rangle$) and using the fact that the kinetic energy is non-negative, we have:
\begin{equation}
\label{eq:apge}
\langle n,N-n|\hat{H}| n, N-n\rangle \ge E_0(n) +  E_0(N-n)
\end{equation}

\subsection{\label{sub:appbound}Bounds on the energy in the limit $N\to \infty$}
{Using the notation introduced in Eq.~(\ref{eq:NOON}), we can write} the total Hartree-product wave function at time $t^*$ as:
\begin{equation}
\label{eq:csum}
|\psi(\underline{x},t^*)\rangle = \sum_{n=0}^N c_n| n, N-n\rangle,
\end{equation}
{which} yields:
\begin{equation}
\langle\psi(\underline{x},t^*)|\hat{H}|\psi(\underline{x},t^*)\rangle = \sum_{n=0}^N |c_n|^2\langle n, N-n|\hat{H} |n, N-n\rangle\;.
\end{equation}
For Hartree-product states, the  $|c_n|^2$ are given by the Binomial distribution~\cite[3.2.2]{HarocheRaimond06}:
\begin{equation}
\label{eq:choose}
|c_n|^2 = \binom{N}{n} p^n (1-p)^{N-n},
\end{equation}
where $p$ is given by Eq.~(\ref{eq:pdeff}).
This distribution is strongly peaked around
\begin{equation}
\overline{n}=Np
\end{equation} 
with root-mean-square deviation 
\begin{equation}
\sigma = \sqrt{Np(1-p)}\;.
\end{equation}
Using Eq.~(\ref{eq:apge}) we thus have:
\begin{align}
\label{eq:hhhh}
\langle\psi(\underline{x},t^*)|\hat{H}|\psi(\underline{x},t^*)\rangle &\simeq \langle \overline{n}, N-\overline{n}|\hat{H} |\overline{n} , N-\overline{n}\rangle
\nonumber \\
&\ge E_0(\overline{n}) +  E_0(N-\overline{n})
\;;
\end{align}
the larger $N$, the better the approximation in the first line of Eq.~(\ref{eq:hhhh}) becomes.
By replacing Eq.~(\ref{eq:RnN}) with
\begin{align}
R = \frac{\overline{n}}N
\end{align}
(which is the same as $p$), the equations and bound derived in Sec.~\ref{sec:bound} are also valid for Hartree-product states. For the relation of the initial center-of-mass energy and the reflection coefficient we thus reproduce the equation~(\ref{eq:EkinGreater}):
\begin{equation}
\label{eq:EkinGreaterapp}
 \frac{{E}_{\rm kin}}{|E_0(N)|}\ge 3(1-R)R\;.
\end{equation}
This leads again to the statement that the magenta/gray region of Fig.~\ref{fig:GP} is forbidden energetically.

\subsection{\label{sub:diff}Possible differences between Hartree-product states and GPE: energy fluctuations}
The above calculation did, however, use the limit $N\to\infty$ for which the distribution~(\ref{eq:choose}) is very narrow. For finite $N$ and low initial center-of-mass kinetic energy [cf.\ Eq.~(\ref{eq:EkinGreaterapp})], energetically disallowed eigenfunctions like those corresponding to $|N/2,N/2\rangle$ would always be an important contribution to the sum~(\ref{eq:csum}). 
The only way to prevent this (cf.\ footnote on page~\pageref{page:footnote}) is to predict that all particles have to be on one side of the scattering potential (i.e., $R=0$ or  $R=1$). 

Hartree-product states for low particle numbers would thus behave differently from the GPE, which is strictly speaking only valid in the limit~(\ref{eq:limit}). Note that the contribution of energetically disallowed states like $|N/2,N/2\rangle$ vanishes only in the limit $N\to \infty$ for which $\lim_{N\to\infty}|c_{N/2}/c_{\overline{n}}|=0$ (if ${N/2}\ne{\overline{n}}$), although taking this limit first and then apply energy arguments might be a physically relevant approach. 

Thus, by extending this energy argument to higher particle numbers one could construct an example for which a Hartree-product state disagrees in its ``digital'' prediction for the reflection coefficient with the GPE. This would, however, not confute the GPE. The fact that Hartree-product states are often used to derive the GPE (cf.~\cite{PethickSmith08}) does not necessarily imply that both always are equivalent; in general, the solutions of the GPE can always be interpreted as $|\varphi(x,t)|^2$ being the single-particle density~\cite{LiebEtAl00} rather than automatically being part of the Hartree-product state~(\ref{eq:HP}).

\end{appendix}


\begin{thebibliography}{10}%
\makeatletter
\providecommand \@ifxundefined [1]{%
 \ifx #1\undefined \expandafter \@firstoftwo
 \else \expandafter \@secondoftwo
\fi
}%
\providecommand \@ifnum [1]{%
 \ifnum #1\expandafter \@firstoftwo
 \else \expandafter \@secondoftwo
\fi
}%
\providecommand \enquote [1]{``#1''}%
\providecommand \bibnamefont  [1]{#1}%
\providecommand \bibfnamefont [1]{#1}%
\providecommand \citenamefont [1]{#1}%
\providecommand\href[0]{\@sanitize\@href}%
\providecommand\@href[1]{\endgroup\@@startlink{#1}\endgroup\@@href}%
\providecommand\@@href[1]{#1\@@endlink}%
\providecommand \@sanitize [0]{\begingroup\catcode`\&12\catcode`\#12\relax}%
\@ifxundefined \pdfoutput {\@firstoftwo}{%
 \@ifnum{\z@=\pdfoutput}{\@firstoftwo}{\@secondoftwo}%
}{%
 \providecommand\@@startlink[1]{\leavevmode}%
 \providecommand\@@endlink[0]{}%
}{%
 \providecommand\@@startlink[1]{%
  \leavevmode
  \pdfstartlink
   attr{/Border[0 0 1 ]/H/I/C[0 1 1]}%
   user{/Subtype/Link/A<</Type/Action/S/URI/URI(#1)>>}%
  \relax
 }%
 \providecommand\@@endlink[0]{\pdfendlink}%
}%
\providecommand \url  [0]{\begingroup\@sanitize \@url }%
\providecommand \@url [1]{\endgroup\@href {#1}{\urlprefix}}%
\providecommand \urlprefix [0]{URL }%
\providecommand \Eprint[0]{\href }%
\@ifxundefined \urlstyle {%
  \providecommand \doi [1]{doi:\discretionary{}{}{}#1}%
}{%
  \providecommand \doi [0]{doi:\discretionary{}{}{}\begingroup
  \urlstyle{rm}\Url }%
}%
\providecommand \doibase [0]{http://dx.doi.org/}%
\providecommand \Doi[1]{\href{\doibase#1}}%
\providecommand \bibAnnote [3]{%
  \BibitemShut{#1}%
  \begin{quotation}\noindent
    \textsc{Key:}\ #2\\\textsc{Annotation:}\ #3%
  \end{quotation}%
}%
\providecommand \bibAnnoteFile [2]{%
  \IfFileExists{#2}{\bibAnnote {#1} {#2} {\input{#2}}}{}%
}%
\providecommand \typeout [0]{\immediate \write \m@ne }%
\providecommand \selectlanguage [0]{\@gobble}%
\providecommand \bibinfo [0]{\@secondoftwo}%
\providecommand \bibfield [0]{\@secondoftwo}%
\providecommand \translation [1]{[#1]}%
\providecommand \BibitemOpen[0]{}%
\providecommand \bibitemStop [0]{}%
\providecommand \bibitemNoStop [0]{.\EOS\space}%
\providecommand \EOS [0]{\spacefactor3000\relax}%
\providecommand \BibitemShut [1]{\csname bibitem#1\endcsname}%
\bibitem{KhaykovichEtAl02}%
  \BibitemOpen
  \bibfield{author}{%
  \bibinfo {author} {\bibfnamefont{L.}~\bibnamefont{Khaykovich}}, \bibinfo
  {author} {\bibfnamefont{F.}~\bibnamefont{Schreck}}, \bibinfo {author}
  {\bibfnamefont{G.}~\bibnamefont{Ferrari}}, \bibinfo {author}
  {\bibfnamefont{T.}~\bibnamefont{Bourdel}}, \bibinfo {author}
  {\bibfnamefont{J.}~\bibnamefont{Cubizolles}}, \bibinfo {author}
  {\bibfnamefont{L.~D.}\ \bibnamefont{Carr}}, \bibinfo {author}
  {\bibfnamefont{Y.}~\bibnamefont{Castin}},\ and\ \bibinfo {author}
  {\bibfnamefont{C.}~\bibnamefont{Salomon}},\ }%
  \bibfield{journal}{%
  \Doi{10.1126/science.1071021}{\bibinfo {journal} {Science}}\ }%
  \textbf{\bibinfo {volume} {296}},\ \bibinfo {pages} {1290} (\bibinfo {year}
  {2002})%
  \bibAnnoteFile{NoStop}{KhaykovichEtAl02}%
\bibitem{StreckerEtAl02}%
  \BibitemOpen
  \bibfield{author}{%
  \bibinfo {author} {\bibfnamefont{K.~E.}\ \bibnamefont{Strecker}}, \bibinfo
  {author} {\bibfnamefont{G.~B.}\ \bibnamefont{Partridge}}, \bibinfo {author}
  {\bibfnamefont{A.~G.}\ \bibnamefont{Truscott}},\ and\ \bibinfo {author}
  {\bibfnamefont{R.~G.}\ \bibnamefont{Hulet}},\ }%
  \bibfield{journal}{%
  \Doi{10.1038/nature747}{\bibinfo {journal} {Nature (London)}}\ }%
  \textbf{\bibinfo {volume} {417}},\ \bibinfo {pages} {150} (\bibinfo {year}
  {2002})%
  \bibAnnoteFile{NoStop}{StreckerEtAl02}%
\bibitem{CornishEtAl06}%
  \BibitemOpen
  \bibfield{author}{%
  \bibinfo {author} {\bibfnamefont{S.~L.}\ \bibnamefont{Cornish}}, \bibinfo
  {author} {\bibfnamefont{S.~T.}\ \bibnamefont{Thompson}},\ and\ \bibinfo
  {author} {\bibfnamefont{C.~E.}\ \bibnamefont{Wieman}},\ }%
  \bibfield{journal}{%
  \Doi{10.1103/PhysRevLett.96.170401}{\bibinfo {journal} {Phys. Rev. Lett.}}\
  }%
  \textbf{\bibinfo {volume} {96}},\ \bibinfo {pages} {170401} (\bibinfo {year}
  {2006})%
  \bibAnnoteFile{NoStop}{CornishEtAl06}%
\bibitem{PethickSmith08}%
  \BibitemOpen
  \bibfield{author}{%
  \bibinfo {author} {\bibfnamefont{C.~J.}\ \bibnamefont{Pethick}}\ and\
  \bibinfo {author} {\bibfnamefont{H.}~\bibnamefont{Smith}},\ }%
  \emph{\bibinfo {title} {Bose-Einstein Condensation in Dilute Gases}}\
  (\bibinfo {publisher} {Cambridge University Press},\ \bibinfo {address}
  {Cambridge},\ \bibinfo {year} {2008})%
  \bibAnnoteFile{NoStop}{PethickSmith08}%
\bibitem{MazetsKurizki06}%
  \BibitemOpen
  \bibfield{author}{%
  \bibinfo {author} {\bibfnamefont{I.~E.}\ \bibnamefont{Mazets}}\ and\ \bibinfo
  {author} {\bibfnamefont{G.}~\bibnamefont{Kurizki}},\ }%
  \bibfield{journal}{%
  \Doi{10.1209/epl/i2006-10260-0}{\bibinfo {journal} {Europhys.\ Lett.}}\ }%
  \textbf{\bibinfo {volume} {76}},\ \bibinfo {pages} {196} (\bibinfo {year}
  {2006})%
  \bibAnnoteFile{NoStop}{MazetsKurizki06}%
\bibitem{WeissCastin09}%
  \BibitemOpen
  \bibfield{author}{%
  \bibinfo {author} {\bibfnamefont{C.}~\bibnamefont{Weiss}}\ and\ \bibinfo
  {author} {\bibfnamefont{Y.}~\bibnamefont{Castin}},\ }%
  \bibfield{journal}{%
  \Doi{10.1103/PhysRevLett.102.010403}{\bibinfo {journal} {Phys.\ Rev.\
  Lett.}}\ }%
  \textbf{\bibinfo {volume} {102}},\ \bibinfo {pages} {010403} (\bibinfo {year}
  {2009})%
  \bibAnnoteFile{NoStop}{WeissCastin09}%
\bibitem{Streltsov09b}%
  \BibitemOpen
  \bibfield{author}{%
  \bibinfo {author} {\bibfnamefont{A.~I.}\ \bibnamefont{Streltsov}}, \bibinfo
  {author} {\bibfnamefont{O.~E.}\ \bibnamefont{Alon}},\ and\ \bibinfo {author}
  {\bibfnamefont{L.~S.}\ \bibnamefont{Cederbaum}},\ }%
  \bibfield{journal}{%
  \Doi{10.1103/PhysRevA.80.043616}{\bibinfo {journal} {Phys.\ Rev.\ A}}\ }%
  \textbf{\bibinfo {volume} {80}},\ \bibinfo {eid} {043616} (\bibinfo {year}
  {2009})%
  \bibAnnoteFile{NoStop}{Streltsov09b}%
\bibitem{CarrEtAl00b}%
  \BibitemOpen
  \bibfield{author}{%
  \bibinfo {author} {\bibfnamefont{L.~D.}\ \bibnamefont{Carr}}, \bibinfo
  {author} {\bibfnamefont{C.~W.}\ \bibnamefont{Clark}},\ and\ \bibinfo {author}
  {\bibfnamefont{W.~P.}\ \bibnamefont{Reinhardt}},\ }%
  \bibfield{journal}{%
  \Doi{10.1103/PhysRevA.62.063611}{\bibinfo {journal} {Phys. Rev. A}}\ }%
  \textbf{\bibinfo {volume} {62}},\ \bibinfo {pages} {063611} (\bibinfo {year}
  {2000})%
  \bibAnnoteFile{NoStop}{CarrEtAl00b}%
\bibitem{AlKhawajaEtAl02}%
  \BibitemOpen
  \bibfield{author}{%
  \bibinfo {author} {\bibfnamefont{U.}~\bibnamefont{Al~Khawaja}}, \bibinfo
  {author} {\bibfnamefont{H.~T.~C.}\ \bibnamefont{Stoof}}, \bibinfo {author}
  {\bibfnamefont{R.~G.}\ \bibnamefont{Hulet}}, \bibinfo {author}
  {\bibfnamefont{K.~E.}\ \bibnamefont{Strecker}},\ and\ \bibinfo {author}
  {\bibfnamefont{G.~B.}\ \bibnamefont{Partridge}},\ }%
  \bibfield{journal}{%
  \Doi{10.1103/PhysRevLett.89.200404}{\bibinfo {journal} {Phys. Rev. Lett.}}\
  }%
  \textbf{\bibinfo {volume} {89}},\ \bibinfo {pages} {200404} (\bibinfo {year}
  {2002})%
  \bibAnnoteFile{NoStop}{AlKhawajaEtAl02}%
\bibitem{StreltsovEtAl11}%
  \BibitemOpen
  \bibfield{author}{%
  \bibinfo {author} {\bibfnamefont{A.~I.}\ \bibnamefont{Streltsov}}, \bibinfo
  {author} {\bibfnamefont{O.~E.}\ \bibnamefont{Alon}},\ and\ \bibinfo {author}
  {\bibfnamefont{L.~S.}\ \bibnamefont{Cederbaum}},\ }%
  \bibfield{journal}{%
  \Doi{10.1103/PhysRevLett.106.240401}{\bibinfo {journal} {Phys. Rev. Lett.}}\
  }%
  \textbf{\bibinfo {volume} {106}},\ \bibinfo {pages} {240401} (\bibinfo {year}
  {2011})%
  \bibAnnoteFile{NoStop}{StreltsovEtAl11}%
\bibitem{SalasnichEtAl02}%
  \BibitemOpen
  \bibfield{author}{%
  \bibinfo {author} {\bibfnamefont{L.}~\bibnamefont{Salasnich}}, \bibinfo
  {author} {\bibfnamefont{A.}~\bibnamefont{Parola}},\ and\ \bibinfo {author}
  {\bibfnamefont{L.}~\bibnamefont{Reatto}},\ }%
  \bibfield{journal}{%
  \Doi{10.1103/PhysRevA.66.043603}{\bibinfo {journal} {Phys. Rev. A}}\ }%
  \textbf{\bibinfo {volume} {66}},\ \bibinfo {pages} {043603} (\bibinfo {year}
  {2002})%
  \bibAnnoteFile{NoStop}{SalasnichEtAl02}%
\bibitem{BuljanEtAl05}%
  \BibitemOpen
  \bibfield{author}{%
  \bibinfo {author} {\bibfnamefont{H.}~\bibnamefont{Buljan}}, \bibinfo {author}
  {\bibfnamefont{M.}~\bibnamefont{Segev}},\ and\ \bibinfo {author}
  {\bibfnamefont{A.}~\bibnamefont{Vardi}},\ }%
  \bibfield{journal}{%
  \Doi{10.1103/PhysRevLett.95.180401}{\bibinfo {journal} {Phys. Rev. Lett.}}\
  }%
  \textbf{\bibinfo {volume} {95}},\ \bibinfo {pages} {180401} (\bibinfo {year}
  {2005})%
  \bibAnnoteFile{NoStop}{BuljanEtAl05}%
\bibitem{KhaykovichMalomed06}%
  \BibitemOpen
  \bibfield{author}{%
  \bibinfo {author} {\bibfnamefont{L.}~\bibnamefont{Khaykovich}}\ and\ \bibinfo
  {author} {\bibfnamefont{B.~A.}\ \bibnamefont{Malomed}},\ }%
  \bibfield{journal}{%
  \Doi{10.1103/PhysRevA.74.023607}{\bibinfo {journal} {Phys. Rev. A}}\ }%
  \textbf{\bibinfo {volume} {74}},\ \bibinfo {pages} {023607} (\bibinfo {year}
  {2006})%
  \bibAnnoteFile{NoStop}{KhaykovichMalomed06}%
\bibitem{MartinEtAl07}%
  \BibitemOpen
  \bibfield{author}{%
  \bibinfo {author} {\bibfnamefont{A.~D.}\ \bibnamefont{Martin}}, \bibinfo
  {author} {\bibfnamefont{C.~S.}\ \bibnamefont{Adams}},\ and\ \bibinfo {author}
  {\bibfnamefont{S.~A.}\ \bibnamefont{Gardiner}},\ }%
  \bibfield{journal}{%
  \Doi{10.1103/PhysRevLett.98.020402}{\bibinfo {journal} {Phys.\ Rev.\ Lett.}}\
  }%
  \textbf{\bibinfo {volume} {98}},\ \bibinfo {eid} {020402} (\bibinfo {year}
  {2007})%
  \bibAnnoteFile{NoStop}{MartinEtAl07}%
\bibitem{StreltsovEtAl08}%
  \BibitemOpen
  \bibfield{author}{%
  \bibinfo {author} {\bibfnamefont{A.~I.}\ \bibnamefont{Streltsov}}, \bibinfo
  {author} {\bibfnamefont{O.~E.}\ \bibnamefont{Alon}},\ and\ \bibinfo {author}
  {\bibfnamefont{L.~S.}\ \bibnamefont{Cederbaum}},\ }%
  \bibfield{journal}{%
  \Doi{10.1103/PhysRevLett.100.130401}{\bibinfo {journal} {Phys.\ Rev.\
  Lett.}}\ }%
  \textbf{\bibinfo {volume} {100}},\ \bibinfo {pages} {130401} (\bibinfo {year}
  {2008})%
  \bibAnnoteFile{NoStop}{StreltsovEtAl08}%
\bibitem{YanayEtAl09}%
  \BibitemOpen
  \bibfield{author}{%
  \bibinfo {author} {\bibfnamefont{H.}~\bibnamefont{Yanay}}, \bibinfo {author}
  {\bibfnamefont{L.}~\bibnamefont{Khaykovich}},\ and\ \bibinfo {author}
  {\bibfnamefont{B.~A.}\ \bibnamefont{Malomed}},\ }%
  \bibfield{journal}{%
  \Doi{10.1063/1.3238246}{\bibinfo {journal} {Chaos}}\ }%
  \textbf{\bibinfo {volume} {19}},\ \bibinfo {pages} {033145} (\bibinfo {year}
  {2009})%
  \bibAnnoteFile{NoStop}{YanayEtAl09}%
\bibitem{CornishEtAl09}%
  \BibitemOpen
  \bibfield{author}{%
  \bibinfo {author} {\bibfnamefont{S.}~\bibnamefont{Cornish}}, \bibinfo
  {author} {\bibfnamefont{N.}~\bibnamefont{Parker}}, \bibinfo {author}
  {\bibfnamefont{A.}~\bibnamefont{Martin}}, \bibinfo {author}
  {\bibfnamefont{T.}~\bibnamefont{Judd}}, \bibinfo {author}
  {\bibfnamefont{R.}~\bibnamefont{Scott}}, \bibinfo {author}
  {\bibfnamefont{T.}~\bibnamefont{Fromhold}},\ and\ \bibinfo {author}
  {\bibfnamefont{C.}~\bibnamefont{Adams}},\ }%
  \bibfield{journal}{%
  \Doi{10.1016/j.physd.2008.07.011}{\bibinfo {journal} {Physica D: Nonlinear
  Phenomena}}\ }%
  \textbf{\bibinfo {volume} {238}},\ \bibinfo {pages} {1299} (\bibinfo {year}
  {2009})%
  \bibAnnoteFile{NoStop}{CornishEtAl09}%
\bibitem{KivsharEtAl90}%
  \BibitemOpen
  \bibfield{author}{%
  \bibinfo {author} {\bibfnamefont{Y.~S.}\ \bibnamefont{Kivshar}}, \bibinfo
  {author} {\bibfnamefont{S.~A.}\ \bibnamefont{Gredeskul}}, \bibinfo {author}
  {\bibfnamefont{A.}~\bibnamefont{S{\'a}nchez}},\ and\ \bibinfo {author}
  {\bibfnamefont{L.}~\bibnamefont{V{\'a}zquez}},\ }%
  \bibfield{journal}{%
  \Doi{10.1103/PhysRevLett.64.1693}{\bibinfo {journal} {Phys. Rev. Lett.}}\ }%
  \textbf{\bibinfo {volume} {64}},\ \bibinfo {pages} {1693} (\bibinfo {year}
  {1990})%
  \bibAnnoteFile{NoStop}{KivsharEtAl90}%
\bibitem{Muller11}%
  \BibitemOpen
  \bibfield{author}{%
  \bibinfo {author} {\bibfnamefont{C.}~\bibnamefont{M{\"u}ller}},\ }%
  \bibfield{journal}{%
  \Doi{10.1007/s00340-011-4425-3}{\bibinfo {journal} {Appl. Phys. B-Lasers O}}\
  }%
  \textbf{\bibinfo {volume} {102}},\ \bibinfo {pages} {459} (\bibinfo {year}
  {2011})%
  \bibAnnoteFile{NoStop}{Muller11}%
\bibitem{Khawaja09}%
  \BibitemOpen
  \bibfield{author}{%
  \bibinfo {author} {\bibfnamefont{U.~A.}\ \bibnamefont{Khawaja}},\ }%
  \bibfield{journal}{%
  \bibinfo {journal} {J. Phys. A}\ }%
  \textbf{\bibinfo {volume} {42}},\ \bibinfo {pages} {265206} (\bibinfo {year}
  {2009})%
  \bibAnnoteFile{NoStop}{Khawaja09}%
\bibitem{ErnstBrand10}%
  \BibitemOpen
  \bibfield{author}{%
  \bibinfo {author} {\bibfnamefont{T.}~\bibnamefont{Ernst}}\ and\ \bibinfo
  {author} {\bibfnamefont{J.}~\bibnamefont{Brand}},\ }%
  \bibfield{journal}{%
  \Doi{10.1103/PhysRevA.81.033614}{\bibinfo {journal} {Phys. Rev. A}}\ }%
  \textbf{\bibinfo {volume} {81}},\ \bibinfo {pages} {033614} (\bibinfo {year}
  {2010})%
  \bibAnnoteFile{NoStop}{ErnstBrand10}%
\bibitem{BillamEtAl11}%
  \BibitemOpen
  \bibfield{author}{%
  \bibinfo {author} {\bibfnamefont{T.~P.}\ \bibnamefont{Billam}}, \bibinfo
  {author} {\bibfnamefont{S.~L.}\ \bibnamefont{Cornish}},\ and\ \bibinfo
  {author} {\bibfnamefont{S.~A.}\ \bibnamefont{Gardiner}},\ }%
  \bibfield{journal}{%
  \Doi{10.1103/PhysRevA.83.041602}{\bibinfo {journal} {Phys. Rev. A}}\ }%
  \textbf{\bibinfo {volume} {83}},\ \bibinfo {pages} {041602(R)} (\bibinfo
  {year} {2011})%
  \bibAnnoteFile{NoStop}{BillamEtAl11}%
\bibitem{MartinRuostekoski2012}%
  \BibitemOpen
  \bibfield{author}{%
  \bibinfo {author} {\bibfnamefont{A.~D.}\ \bibnamefont{Martin}}\ and\ \bibinfo
  {author} {\bibfnamefont{J.}~\bibnamefont{Ruostekoski}},\ }%
  \bibfield{journal}{%
  \Doi{10.1088/1367-2630/14/4/043040}{\bibinfo {journal} {New J. Phys.}}\ }%
  \textbf{\bibinfo {volume} {14}},\ \bibinfo {pages} {043040} (\bibinfo {year}
  {2012})%
  \bibAnnoteFile{NoStop}{MartinRuostekoski2012}%
\bibitem{HelmEtAl2012}%
  \BibitemOpen
  \bibfield{author}{%
  \bibinfo {author} {\bibfnamefont{J.~L.}\ \bibnamefont{Helm}}, \bibinfo
  {author} {\bibfnamefont{T.~P.}\ \bibnamefont{Billam}},\ and\ \bibinfo
  {author} {\bibfnamefont{S.~A.}\ \bibnamefont{Gardiner}},\ }%
  \bibfield{journal}{%
  \Doi{10.1103/PhysRevA.85.053621}{\bibinfo {journal} {Phys. Rev. A}}\ }%
  \textbf{\bibinfo {volume} {85}},\ \bibinfo {pages} {053621} (\bibinfo {year}
  {2012})%
  \bibAnnoteFile{NoStop}{HelmEtAl2012}%
\bibitem{LeeBrand06}%
  \BibitemOpen
  \bibfield{author}{%
  \bibinfo {author} {\bibfnamefont{C.}~\bibnamefont{Lee}}\ and\ \bibinfo
  {author} {\bibfnamefont{J.}~\bibnamefont{Brand}},\ }%
  \bibfield{journal}{%
  \Doi{10.1209/epl/i2005-10408-4}{\bibinfo {journal} {Europhys. Lett.}}\ }%
  \textbf{\bibinfo {volume} {73}},\ \bibinfo {pages} {321} (\bibinfo {year}
  {2006})%
  \bibAnnoteFile{NoStop}{LeeBrand06}%
\bibitem{Hulet10}%
  \BibitemOpen
  \bibinfo {note} {R.\ G.\ Hulet, private communication}%
  \bibAnnoteFile{NoStop}{Hulet10}%
\bibitem{Hulet10b}%
  \BibitemOpen
  \bibfield{author}{%
  \bibinfo {author} {\bibfnamefont{S.~E.}\ \bibnamefont{Pollack}}, \bibinfo
  {author} {\bibfnamefont{D.}~\bibnamefont{Dries}}, \bibinfo {author}
  {\bibfnamefont{E.~J.}\ \bibnamefont{Olson}},\ and\ \bibinfo {author}
  {\bibfnamefont{R.~G.}\ \bibnamefont{Hulet}},\ }%
  \bibinfo {note} {2010 DAMOP: Conference abstract,
  http://meetings.aps.org/link/BAPS.2010.DAMOP.R4.1}%
  \bibAnnoteFile{NoStop}{Hulet10b}%
\bibitem{LiebLiniger63}%
  \BibitemOpen
  \bibfield{author}{%
  \bibinfo {author} {\bibfnamefont{E.~H.}\ \bibnamefont{Lieb}}\ and\ \bibinfo
  {author} {\bibfnamefont{W.}~\bibnamefont{Liniger}},\ }%
  \bibfield{journal}{%
  \Doi{10.1103/PhysRev.130.1605}{\bibinfo {journal} {Phys. Rev.}}\ }%
  \textbf{\bibinfo {volume} {130}},\ \bibinfo {pages} {1605} (\bibinfo {year}
  {1963})%
  \bibAnnoteFile{NoStop}{LiebLiniger63}%
\bibitem{McGuire64}%
  \BibitemOpen
  \bibfield{author}{%
  \bibinfo {author} {\bibfnamefont{J.~B.}\ \bibnamefont{McGuire}},\ }%
  \bibfield{journal}{%
  \Doi{10.1063/1.1704156}{\bibinfo {journal} {J. Math. Phys.}}\ }%
  \textbf{\bibinfo {volume} {5}},\ \bibinfo {pages} {622} (\bibinfo {year}
  {1964})%
  \bibAnnoteFile{NoStop}{McGuire64}%
\bibitem{LiebEtAl00}%
  \BibitemOpen
  \bibfield{author}{%
  \bibinfo {author} {\bibfnamefont{E.~H.}\ \bibnamefont{Lieb}}, \bibinfo
  {author} {\bibfnamefont{R.}~\bibnamefont{Seiringer}},\ and\ \bibinfo {author}
  {\bibfnamefont{J.}~\bibnamefont{Yngvason}},\ }%
  \bibfield{journal}{%
  \Doi{10.1103/PhysRevA.61.043602}{\bibinfo {journal} {Phys. Rev. A}}\ }%
  \textbf{\bibinfo {volume} {61}},\ \bibinfo {pages} {043602} (\bibinfo {year}
  {2000})%
  \bibAnnoteFile{NoStop}{LiebEtAl00}%
\bibitem{CastinHerzog01}%
  \BibitemOpen
  \bibfield{author}{%
  \bibinfo {author} {\bibfnamefont{Y.}~\bibnamefont{Castin}}\ and\ \bibinfo
  {author} {\bibfnamefont{C.}~\bibnamefont{Herzog}},\ }%
  \bibfield{journal}{%
  \Doi{10.1016/S1296-2147(01)01183-0}{\bibinfo {journal} {C. R. Acad. Sci.
  Paris, Ser. 4}}\ }%
  \textbf{\bibinfo {volume} {2}},\ \bibinfo {pages} {419} (\bibinfo {year}
  {2001})%
  \bibAnnoteFile{NoStop}{CastinHerzog01}%
\bibitem{CalogeroDegasperis75}%
  \BibitemOpen
  \bibfield{author}{%
  \bibinfo {author} {\bibfnamefont{F.}~\bibnamefont{Calogero}}\ and\ \bibinfo
  {author} {\bibfnamefont{A.}~\bibnamefont{Degasperis}},\ }%
  \bibfield{journal}{%
  \Doi{10.1103/PhysRevA.11.265}{\bibinfo {journal} {Phys. Rev. A}}\ }%
  \textbf{\bibinfo {volume} {11}},\ \bibinfo {pages} {265} (\bibinfo {year}
  {1975})%
  \bibAnnoteFile{NoStop}{CalogeroDegasperis75}%
\bibitem{CastinDalibard97}%
  \BibitemOpen
  \bibfield{author}{%
  \bibinfo {author} {\bibfnamefont{Y.}~\bibnamefont{Castin}}\ and\ \bibinfo
  {author} {\bibfnamefont{J.}~\bibnamefont{Dalibard}},\ }%
  \bibfield{journal}{%
  \Doi{10.1103/PhysRevA.55.4330}{\bibinfo {journal} {Phys. Rev. A}}\ }%
  \textbf{\bibinfo {volume} {55}},\ \bibinfo {pages} {4330} (\bibinfo {year}
  {1997})%
  \bibAnnoteFile{NoStop}{CastinDalibard97}%
\bibitem{RuostekoskiEtAl98}%
  \BibitemOpen
  \bibfield{author}{%
  \bibinfo {author} {\bibfnamefont{J.}~\bibnamefont{Ruostekoski}}, \bibinfo
  {author} {\bibfnamefont{M.~J.}\ \bibnamefont{Collett}}, \bibinfo {author}
  {\bibfnamefont{R.}~\bibnamefont{Graham}},\ and\ \bibinfo {author}
  {\bibfnamefont{D.~F.}\ \bibnamefont{Walls}},\ }%
  \bibfield{journal}{%
  \Doi{10.1103/PhysRevA.57.511}{\bibinfo {journal} {Phys. Rev. A}}\ }%
  \textbf{\bibinfo {volume} {57}},\ \bibinfo {pages} {511} (\bibinfo {year}
  {1998})%
  \bibAnnoteFile{NoStop}{RuostekoskiEtAl98}%
\bibitem{DunninghamBurnett01}%
  \BibitemOpen
  \bibfield{author}{%
  \bibinfo {author} {\bibfnamefont{J.~A.}\ \bibnamefont{Dunningham}}\ and\
  \bibinfo {author} {\bibfnamefont{K.}~\bibnamefont{Burnett}},\ }%
  \bibfield{journal}{%
  \Doi{10.1080/09500340108240890}{\bibinfo {journal} {J.\ Mod.\ Opt.}}\ }%
  \textbf{\bibinfo {volume} {48}},\ \bibinfo {pages} {1837} (\bibinfo {year}
  {2001})%
  \bibAnnoteFile{NoStop}{DunninghamBurnett01}%
\bibitem{MicheliEtAl03}%
  \BibitemOpen
  \bibfield{author}{%
  \bibinfo {author} {\bibfnamefont{A.}~\bibnamefont{Micheli}}, \bibinfo
  {author} {\bibfnamefont{D.}~\bibnamefont{Jaksch}}, \bibinfo {author}
  {\bibfnamefont{J.~I.}\ \bibnamefont{Cirac}},\ and\ \bibinfo {author}
  {\bibfnamefont{P.}~\bibnamefont{Zoller}},\ }%
  \bibfield{journal}{%
  \Doi{10.1103/PhysRevA.67.013607}{\bibinfo {journal} {Phys. Rev. A}}\ }%
  \textbf{\bibinfo {volume} {67}},\ \bibinfo {pages} {013607} (\bibinfo {year}
  {2003})%
  \bibAnnoteFile{NoStop}{MicheliEtAl03}%
\bibitem{MahmudEtAl05}%
  \BibitemOpen
  \bibfield{author}{%
  \bibinfo {author} {\bibfnamefont{K.~W.}\ \bibnamefont{Mahmud}}, \bibinfo
  {author} {\bibfnamefont{H.}~\bibnamefont{Perry}},\ and\ \bibinfo {author}
  {\bibfnamefont{W.~P.}\ \bibnamefont{Reinhardt}},\ }%
  \bibfield{journal}{%
  \Doi{10.1103/PhysRevA.71.023615}{\bibinfo {journal} {Phys. Rev. A}}\ }%
  \textbf{\bibinfo {volume} {71}},\ \bibinfo {pages} {023615} (\bibinfo {year}
  {2005})%
  \bibAnnoteFile{NoStop}{MahmudEtAl05}%
\bibitem{Dounas-frazerEtAl07}%
  \BibitemOpen
  \bibfield{author}{%
  \bibinfo {author} {\bibfnamefont{D.~R.}\ \bibnamefont{Dounas-Frazer}},
  \bibinfo {author} {\bibfnamefont{A.~M.}\ \bibnamefont{Hermundstad}},\ and\
  \bibinfo {author} {\bibfnamefont{L.~D.}\ \bibnamefont{Carr}},\ }%
  \bibfield{journal}{%
  \Doi{10.1103/PhysRevLett.99.200402}{\bibinfo {journal} {Phys.\ Rev.\ Lett.}}\
  }%
  \textbf{\bibinfo {volume} {99}},\ \bibinfo {eid} {200402} (\bibinfo {year}
  {2007})%
  \bibAnnoteFile{NoStop}{Dounas-frazerEtAl07}%
\bibitem{DagninoEtAl09}%
  \BibitemOpen
  \bibfield{author}{%
  \bibinfo {author} {\bibfnamefont{D.}~\bibnamefont{Dagnino}}, \bibinfo
  {author} {\bibfnamefont{N.}~\bibnamefont{Barberan}}, \bibinfo {author}
  {\bibfnamefont{M.}~\bibnamefont{Lewenstein}},\ and\ \bibinfo {author}
  {\bibfnamefont{J.}~\bibnamefont{Dalibard}},\ }%
  \bibfield{journal}{%
  \Doi{10.1038/nphys1277}{\bibinfo {journal} {Nature Phys.}}\ }%
  \textbf{\bibinfo {volume} {5}},\ \bibinfo {pages} {431} (\bibinfo {year}
  {2009})%
  \bibAnnoteFile{NoStop}{DagninoEtAl09}%
\bibitem{GertjerenkenEtAl10}%
  \BibitemOpen
  \bibfield{author}{%
  \bibinfo {author} {\bibfnamefont{B.}~\bibnamefont{Gertjerenken}}, \bibinfo
  {author} {\bibfnamefont{S.}~\bibnamefont{Arlinghaus}}, \bibinfo {author}
  {\bibfnamefont{N.}~\bibnamefont{Teichmann}},\ and\ \bibinfo {author}
  {\bibfnamefont{C.}~\bibnamefont{Weiss}},\ }%
  \bibfield{journal}{%
  \Doi{10.1103/PhysRevA.82.023620}{\bibinfo {journal} {Phys. Rev. A}}\ }%
  \textbf{\bibinfo {volume} {82}},\ \bibinfo {pages} {023620} (\bibinfo {year}
  {2010})%
  \bibAnnoteFile{NoStop}{GertjerenkenEtAl10}%
\bibitem{GarciaMarchEtAl11}%
  \BibitemOpen
  \bibfield{author}{%
  \bibinfo {author} {\bibfnamefont{M.~A.}\ \bibnamefont{Garc\'\i{}a-March}},
  \bibinfo {author} {\bibfnamefont{D.~R.}\ \bibnamefont{Dounas-Frazer}},\ and\
  \bibinfo {author} {\bibfnamefont{L.~D.}\ \bibnamefont{Carr}},\ }%
  \bibfield{journal}{%
  \Doi{10.1103/PhysRevA.83.043612}{\bibinfo {journal} {Phys. Rev. A}}\ }%
  \textbf{\bibinfo {volume} {83}},\ \bibinfo {pages} {043612} (\bibinfo {year}
  {2011})%
  \bibAnnoteFile{NoStop}{GarciaMarchEtAl11}%
\bibitem{MazzarellaEtAl11}%
  \BibitemOpen
  \bibfield{author}{%
  \bibinfo {author} {\bibfnamefont{G.}~\bibnamefont{Mazzarella}}, \bibinfo
  {author} {\bibfnamefont{L.}~\bibnamefont{Salasnich}}, \bibinfo {author}
  {\bibfnamefont{A.}~\bibnamefont{Parola}},\ and\ \bibinfo {author}
  {\bibfnamefont{F.}~\bibnamefont{Toigo}},\ }%
  \bibfield{journal}{%
  \Doi{10.1103/PhysRevA.83.053607}{\bibinfo {journal} {Phys. Rev. A}}\ }%
  \textbf{\bibinfo {volume} {83}},\ \bibinfo {pages} {053607} (\bibinfo {year}
  {2011})%
  \bibAnnoteFile{NoStop}{MazzarellaEtAl11}%
\bibitem{DellAnna12}%
  \BibitemOpen
  \bibfield{author}{%
  \bibinfo {author} {\bibfnamefont{L.}~\bibnamefont{Dell'Anna}},\ }%
  \bibfield{journal}{%
  \Doi{10.1103/PhysRevA.85.053608}{\bibinfo {journal} {Phys. Rev. A}}\ }%
  \textbf{\bibinfo {volume} {85}},\ \bibinfo {pages} {053608} (\bibinfo {year}
  {2012})%
  \bibAnnoteFile{NoStop}{DellAnna12}%
\bibitem{HolmerEtAl07}%
  \BibitemOpen
  \bibfield{author}{%
  \bibinfo {author} {\bibfnamefont{J.}~\bibnamefont{Holmer}}, \bibinfo {author}
  {\bibfnamefont{J.}~\bibnamefont{Marzuola}},\ and\ \bibinfo {author}
  {\bibfnamefont{M.}~\bibnamefont{Zworski}},\ }%
  \bibfield{journal}{%
  \Doi{10.1007/s00332-006-0807-9}{\bibinfo {journal} {J. Nonlinear Sci.}}\ }%
  \textbf{\bibinfo {volume} {17}},\ \bibinfo {pages} {349} (\bibinfo {year}
  {2007})%
  \bibAnnoteFile{NoStop}{HolmerEtAl07}%
\bibitem{HoldawayEtAl2012}%
  \BibitemOpen
  \bibfield{author}{%
  \bibinfo {author} {\bibfnamefont{D.~I.~H.}\ \bibnamefont{Holdaway}}, \bibinfo
  {author} {\bibfnamefont{C.}~\bibnamefont{Weiss}},\ and\ \bibinfo {author}
  {\bibfnamefont{S.~A.}\ \bibnamefont{Gardiner}},\ }%
  \bibfield{journal}{%
  \Doi{10.1103/PhysRevA.85.053618}{\bibinfo {journal} {Phys. Rev. A}}\ }%
  \textbf{\bibinfo {volume} {85}},\ \bibinfo {pages} {053618} (\bibinfo {year}
  {2012})%
  \bibAnnoteFile{NoStop}{HoldawayEtAl2012}%
\bibitem{SachaEtAl09}%
  \BibitemOpen
  \bibfield{author}{%
  \bibinfo {author} {\bibfnamefont{K.}~\bibnamefont{Sacha}}, \bibinfo {author}
  {\bibfnamefont{C.~A.}\ \bibnamefont{M\"uller}}, \bibinfo {author}
  {\bibfnamefont{D.}~\bibnamefont{Delande}},\ and\ \bibinfo {author}
  {\bibfnamefont{J.}~\bibnamefont{Zakrzewski}},\ }%
  \bibfield{journal}{%
  \Doi{10.1103/PhysRevLett.103.210402}{\bibinfo {journal} {Phys. Rev. Lett.}}\
  }%
  \textbf{\bibinfo {volume} {103}},\ \bibinfo {pages} {210402} (\bibinfo {year}
  {2009})%
  \bibAnnoteFile{NoStop}{SachaEtAl09}%
\bibitem{WeissCastin12}%
  \BibitemOpen
  \bibfield{author}{%
  \bibinfo {author} {\bibfnamefont{C.}~\bibnamefont{{Weiss}}}\ and\ \bibinfo
  {author} {\bibfnamefont{Y.}~\bibnamefont{{Castin}}},\ }%
  \bibfield{journal}{%
  \bibinfo {journal} {ArXiv e-prints}}%
   (\bibinfo {year} {2012}),\
  \Eprint{http://arxiv.org/abs/1207.7131}{arXiv:1207.7131}%
  \bibAnnoteFile{NoStop}{WeissCastin12}%
\bibitem{Castin09}%
  \BibitemOpen
  \bibfield{author}{%
  \bibinfo {author} {\bibfnamefont{Y.}~\bibnamefont{Castin}},\ }%
  \bibfield{journal}{%
  \Doi{10.1140/epjb/e2008-00407-3}{\bibinfo {journal} {Eur. Phys. J. B}}\ }%
  \textbf{\bibinfo {volume} {68}},\ \bibinfo {pages} {317} (\bibinfo {year}
  {2009})%
  \bibAnnoteFile{NoStop}{Castin09}%
\bibitem{Gross11}%
  \BibitemOpen
  \bibinfo {note} {C.\ Gross, private communication}%
  \bibAnnoteFile{NoStop}{Gross11}%
\bibitem{Fluegge90}%
  \BibitemOpen
  \bibfield{author}{%
  \bibinfo {author} {\bibfnamefont{S.}~\bibnamefont{{Fl\"ugge}}},\ }%
  \emph{\bibinfo {title} {Rechenmethoden der Quantentheorie}}\ (\bibinfo
  {publisher} {Springer},\ \bibinfo {address} {Berlin},\ \bibinfo {year}
  {1990})%
  \bibAnnoteFile{NoStop}{Fluegge90}%
\bibitem{DimopoulosEtAl07}%
  \BibitemOpen
  \bibfield{author}{%
  \bibinfo {author} {\bibfnamefont{S.}~\bibnamefont{Dimopoulos}}, \bibinfo
  {author} {\bibfnamefont{P.~W.}\ \bibnamefont{Graham}}, \bibinfo {author}
  {\bibfnamefont{J.~M.}\ \bibnamefont{Hogan}},\ and\ \bibinfo {author}
  {\bibfnamefont{M.~A.}\ \bibnamefont{Kasevich}},\ }%
  \bibfield{journal}{%
  \Doi{10.1103/PhysRevLett.98.111102}{\bibinfo {journal} {Phys. Rev. Lett.}}\
  }%
  \textbf{\bibinfo {volume} {98}},\ \bibinfo {pages} {111102} (\bibinfo {year}
  {2007})%
  \bibAnnoteFile{NoStop}{DimopoulosEtAl07}%
\bibitem{GrossEtAl10}%
  \BibitemOpen
  \bibfield{author}{%
  \bibinfo {author} {\bibfnamefont{C.}~\bibnamefont{Gross}}, \bibinfo {author}
  {\bibfnamefont{T.}~\bibnamefont{Zibold}}, \bibinfo {author}
  {\bibfnamefont{E.}~\bibnamefont{Nicklas}}, \bibinfo {author}
  {\bibfnamefont{J.}~\bibnamefont{Esteve}},\ and\ \bibinfo {author}
  {\bibfnamefont{M.~K.}\ \bibnamefont{Oberthaler}},\ }%
  \bibfield{journal}{%
  \bibinfo {journal} {Nature (London)}\ }%
  \textbf{\bibinfo {volume} {464}},\ \bibinfo {pages} {1165} (\bibinfo {year}
  {2010})%
  \bibAnnoteFile{NoStop}{GrossEtAl10}%
\bibitem{SorensenEtAl12}%
  \BibitemOpen
  \bibfield{author}{%
  \bibinfo {author} {\bibfnamefont{O.~S.}\ \bibnamefont{S\o{}rensen}}, \bibinfo
  {author} {\bibfnamefont{S.}~\bibnamefont{Gammelmark}},\ and\ \bibinfo
  {author} {\bibfnamefont{K.}~\bibnamefont{M\o{}lmer}},\ }%
  \bibfield{journal}{%
  \Doi{10.1103/PhysRevA.85.043617}{\bibinfo {journal} {Phys. Rev. A}}\ }%
  \textbf{\bibinfo {volume} {85}},\ \bibinfo {pages} {043617} (\bibinfo {year}
  {2012})%
  \bibAnnoteFile{NoStop}{SorensenEtAl12}%
\bibitem{GlickCarr11}%
  \BibitemOpen
  \bibfield{author}{%
  \bibinfo {author} {\bibfnamefont{J.~A.}\ \bibnamefont{{Glick}}}\ and\
  \bibinfo {author} {\bibfnamefont{L.~D.}\ \bibnamefont{{Carr}}},\ }%
  \bibfield{journal}{%
  \bibinfo {journal} {ArXiv e-prints}}%
   (\bibinfo {year} {2011}),\
  \Eprint{http://arxiv.org/abs/1105.5164}{arXiv:1105.5164
  [cond-mat.quant-gas]}%
  \bibAnnoteFile{NoStop}{GlickCarr11}%
\bibitem{ShampineGordon75}%
  \BibitemOpen
  \bibfield{author}{%
  \bibinfo {author} {\bibfnamefont{L.~F.}\ \bibnamefont{Shampine}}\ and\
  \bibinfo {author} {\bibfnamefont{M.~K.}\ \bibnamefont{Gordon}},\ }%
  \emph{\bibinfo {title} {Computer Solution of Ordinary Differential
  Equations}}\ (\bibinfo {publisher} {Freeman},\ \bibinfo {address} {San
  Francisco},\ \bibinfo {year} {1975})%
  \bibAnnoteFile{NoStop}{ShampineGordon75}%
\bibitem{boffetta_osborne_jcp_1992}%
  \BibitemOpen
  \bibfield{author}{%
  \bibinfo {author} {\bibfnamefont{G.}~\bibnamefont{Boffetta}}\ and\ \bibinfo
  {author} {\bibfnamefont{A.~R.}\ \bibnamefont{Osborne}},\ }%
  \bibfield{journal}{%
  \Doi{10.1016/0021-9991(92)90370-E}{\bibinfo {journal} {J. Comp. Phys.}}\ }%
  \textbf{\bibinfo {volume} {102}},\ \bibinfo {pages} {252} (\bibinfo {year}
  {1992})%
  \bibAnnoteFile{NoStop}{boffetta_osborne_jcp_1992}%
\bibitem{WangEtAl12}%
  \BibitemOpen
  \bibfield{author}{%
  \bibinfo {author} {\bibfnamefont{C.-H.}\ \bibnamefont{{Wang}}}, \bibinfo
  {author} {\bibfnamefont{T.-M.}\ \bibnamefont{{Hong}}}, \bibinfo {author}
  {\bibfnamefont{R.-K.}\ \bibnamefont{{Lee}}},\ and\ \bibinfo {author}
  {\bibfnamefont{D.-W.}\ \bibnamefont{{Wang}}},\ }%
  \bibfield{journal}{%
  \bibinfo {journal} {ArXiv e-prints}}%
   (\bibinfo {year} {2012}),\
  \Eprint{http://arxiv.org/abs/1206.1606}{arXiv:1206.1606
  [cond-mat.quant-gas]}%
  \bibAnnoteFile{NoStop}{WangEtAl12}%
\bibitem{HarocheRaimond06}%
  \BibitemOpen
  \bibfield{author}{%
  \bibinfo {author} {\bibfnamefont{S.}~\bibnamefont{Haroche}}\ and\ \bibinfo
  {author} {\bibfnamefont{J.-M.}\ \bibnamefont{Raimond}},\ }%
  \emph{\bibinfo {title} {Exploring the Quantum -- Atoms, Cavities and
  Photons}}\ (\bibinfo {publisher} {Oxford University Press},\ \bibinfo
  {address} {Oxford},\ \bibinfo {year} {2006})%
  \bibAnnoteFile{NoStop}{HarocheRaimond06}%
\end{thebibliography}
\end{document}